\documentclass[final,authoryear,5p,times,twocolumn]{elsarticle}

\usepackage{amssymb}

\def\aj{AJ} 
 
\def\araa{ARA\&A} 
\def\apj{ApJ} 
\def\apjl{ApJ} 
\def\apjs{ApJS}

\def\aap{A\&A} 
 
\def\aaps{A\&AS}

\def\mnras{MNRAS}

\def\fcp{Fund.~Cosmic~Phys.}

\def\memsai{Mem.~Soc.~Astron.~Ital.}

\journal{New Astronomy Reviews}

\begin{document}

\begin{frontmatter}

\title{The stochastic nature of stellar population modelling}

\author[iaa,iac,ull]{Miguel Cervi\~no}

\address[iaa]{Instituto de Astrof\'{\i}sica de Andluc\'{\i}a (IAA-CSIC), Placeta de la Astronom\'{\i}a s/n, 18028, Granada, Spain}
\address[iac]{Instituto de Astrof\'{\i}sica de Canarias (IAC), 38205 La Laguna, Tenerife, Spain}
\address[ull]{Departamento de Astrof\'{\i}sica, Universidad de La Laguna (ULL), E-38206 La Laguna, Spain}
\ead{mcs@iac.es}

\begin{abstract}
Since the early 1970s, stellar population modelling has been one of the basic tools for understanding the physics of unresolved systems from observation of their integrated light. Models allow us to relate the integrated spectra (or colours) of a system with the evolutionary status of the stars of which it is composed and hence to infer how the system has evolved from its formation to its present stage. On average, observational data follow model predictions, but with some scatter, so that systems with the same physical parameters (age, metallicity, total mass) produce a variety of integrated spectra. The fewer the stars in a system, the larger is the scatter. Such scatter is sometimes much larger than the observational errors, reflecting its physical nature. This situation has led to the development in recent years (especially since 2010) of Monte Carlo models of stellar populations. Some authors have proposed that such models are more realistic than state-of-the-art standard synthesis codes that produce the mean of the distribution of Monte Carlo models. 

In this review, I show that these two modelling strategies are actually equivalent, and that they are not in opposition to each other. They are just different ways of describing the probability distributions intrinsic in the very modelling of stellar populations. I show the advantages and limitations of each strategy and how they complement each other. I also show the implications of the probabilistic description of stellar populations in the application of models to observational data obtained with high-resolution observational facilities. Finally, I outline some possible developments that could be realized in stellar population modelling in the near future.
\end{abstract}


\end{frontmatter}

\begin{flushright}
\begin{tabular}{r}
{\small You only get a measure of order and control}\\
{\small when you embrace randomness.}\\
{\small (N.N. Taleb, \textit{Antifragile})
}
\end{tabular}
\end{flushright}

\section{Introduction}
\label{}

\subsection{Motivation}
Open your window and take a look at the night sky on a clear night. You can see lots of stars, lots of different types of stars. You work in astrophysics so you can count the stars and measure their light (you can use different photometric filters or measure spectra), sum the individual observations, and obtain the integrated luminosity (magnitudes or spectra). After that, ask some colleagues to do the same experiment, observing the same number of stars, $\cal N$, you have counted in distant places, let us say $n_\mathrm{sam}$ colleagues. In fact, they will see different regions of the sky so they will observe different stars; hence, you are sampling the sky with $n_\mathrm{sam}$ elements, each with $\cal N$ stars. Compare the integrated luminosities and, almost certainly, they will differ. But you are looking in the Sun's neighbourhood where you can define the set of physical conditions that define the stars that you would observe (initial mass function, star formation history, age, metallicity); hence, your results and those of your colleagues should be consistent with these physical conditions, although they differ from each other, right?

Now take your favourite synthesis code, include the physical conditions, and obtain the integrated luminosity, magnitude, and spectra. In addition, perform millions ($n_\mathrm{sam}$ is millions) of Monte Carlo simulations with those physical conditions using $\cal N$ stars in each simulation. Almost certainly, neither the integrated luminosities obtained by the code nor any of those obtained by Monte Carlo simulations equal the ones you or any of your colleagues have obtained. However, such scatter is an inherent result of nature and, as it should be, is implicit in the modelling. Therefore, you see scatter in both observations by you and your colleagues and the Monte Carlo simulations; but where is the scatter in the standard results provided by synthesis codes? After all, most codes only produce a single result for given physical conditions.

After some time thinking about this (several years in my case), you realize that the results of the Monte Carlo simulations are {\it distributed}, and you can glimpse the shape of such distributions. Moreover, you realize that your observations are inside the distribution of the Monte Carlo results once binned. Maybe they are not in the most populated bin, maybe some of them are in a low-frequency bin, but they are inside the distribution of simulations, exactly as if they were additional simulations. You then obtain the mean value of the distribution and you realize that it is suspiciously similar to the value obtained by the synthesis code using $\cal N$ as a scaling factor. It is also applied to the mean value obtained from your observational set. Furthermore, if you obtain the variance (a measure of scatter) for the Monte Carlo set and the observational set and divide them by the mean values you obtained before, the results are similar to each other and to the (so-called) surface brightness fluctuations, SBF, the synthesis code would produce. It is valid for ${\cal N} = 1$ and ${\cal N} = \infty$, although the larger that $n_\mathrm{sam}$ is, the more similar the results are.

From this experiment you realize that the shape of the integrated luminosity distribution changes with $\cal N$ and would actually be power law-like, multimodal, bimodal, or Gaussian-like, but {\it synthesis models, the standard synthesis models you have known for years, always produce the correct mean value of the simulation /observational set. Standard models are also able to obtain a correct measure of the dispersion of possible results (some codes have actually computed it since the late 1980s)}. A different issue is that most standard synthesis models provide only the mean value of the distribution of possible luminosities; hence, they lose part of their predicting power.

If you perform this experiment using different photometric filters or integrated spectra, you also realize that, depending on the wavelength, you obtain different values of the mean {\it and the variance}. You will certainly find that the scatter is greater for red than for blue wavelengths. But this result is consistent with the SBF values obtained from the population synthesis code, which does not include observational errors, so such variation in the scatter with wavelength is a physical result valid for ${\cal N} = 1$ and ${\cal N} = \infty$. This is somewhat confusing since it implies that knowledge of the emission in a spectral region does not directly provide knowledge regarding other spectral regions. There is no perfect correlation between different wavelengths, only partial correlations. You might then think that a $\chi^2$ fitting including only the observational error and not the theoretical scatter might not be very good; it might be better to include the physical dispersion obtained from the models. It would be even better to include the theoretical correlation coefficients among different wavelengths.

You then realize that the meaning of sampling effects are primarily related to $n_\mathrm{sam}$ and only secondarily to $\cal N$, and you begin to think about how to take advantage of this. We know that $n_\mathrm{sam} \times {\cal N}$ is the total number of stars in the system, $N_\mathrm{tot}$; hence, we can establish that the analysis of resolved populations using colour--magnitude diagrams (CMDs) uses ${\cal N} = 1$ and $n_\mathrm{sam} = N_\mathrm{tot}$. In fact, CMDs are the best option for inferring stellar population parameters since you have information about all the stars {\it and about how their luminosities are distributed}. Analysis of the integrated properties of a fully unresolved system uses ${\cal N} = N_\mathrm{tot}$ and $n_\mathrm{sam} = 1$. Systems that are not fully resolved or not fully unresolved are becoming the norm with new and future observational facilities; they have $\cal N$ and $n_\mathrm{sam}$ values in the range 1 and $N_\mathrm{tot}$. The stars are not resolved, but you have more information than only the integrated properties: you have more than one event to sample the distribution of possible luminosities. If you apply this idea to integral field units (IFUs), you realize that IFU observations (the overall set of IFU observations, not the individual observation of an IFU) can be used to sample the distribution of possible luminosities and to obtain more accurate inferences about stellar populations in the system. If you understand how the distributions of integrated luminosities vary with $\cal N$ and the wavelength, going from power-law-like, multimodal, bimodal and Gaussian-like when $\cal N$ increases, you can apply this to any set of systems. For instance, it can be applied to the globular cluster system and to predict $\cal N$, age, metallicity, and star formation history ranges, where you would observe bimodal colour distributions.

Stellar population modelling is intrinsically probabilistic by construction and is independent whether we are aware of it or not. It describes all the possible situations once physical conditions and the number of stars, $\cal N$, are defined. It is actually the most accurate description we have of stellar populations. Mother Nature is intrinsically stochastic, playing with the whole range of available possibilities for given physics of the system. If you realize that, you will also realize that the observed scatter (once corrected for observational errors) contains physical information, and you will look for such scatter.

The study of stochasticity in the modelling of stellar populations is not new. However, in recent years, with the increasing resolution of observational facilities, the subject has become more and more relevant. There are several papers that address partial aspects related to the stochasticity of stellar populations, but almost none that address the issue from a general point of view and exploit the implications. This last point is the objective of this paper. To achieve this, I assume that all synthesis models are equivalent and correct since comparison of synthesis model results is beyond the scope of this paper. I begin with a brief historical outline of the evolution of stellar population modelling. I continue with an analysis of the origin of stochasticity in modelling in Section \ref{sec:origin}. Parametric and Monte Carlo descriptions of stellar population modelling are presented in Sections \ref{sec:param} and \ref{sec:montecarlo}, respectively. I describe the implications of stochasticity in the use of stellar population codes in Section \ref{sec:implications}. This section includes some rules of thumb for the use of synthesis models. I outline an unexplored area in which stochasticity could play a role in Section \ref{sec:open}. My conclusions are presented in Section \ref{sec:conclu}.

\subsection{A short historical review}
\label{sec:histo}

\begin{figure}
\resizebox{\hsize}{!}{\includegraphics[angle=-90]{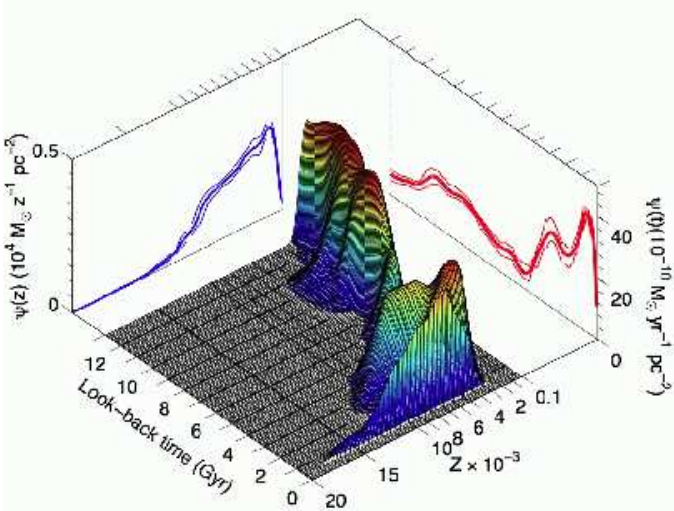}}
\caption[]{Population box for a Large Magellanic Cloud (LMC) field taken from \citet{Mesetal13}. Courtesy of C. Gallart.}
\label{fig:CMDcarma}
\end{figure}

The stellar population concept can be traced back to the work of \citet{Baa44} through the empirical characterization of CMDs in different systems by direct star counting. Similarities and differences in star cluster CMD structure allowed them to be classified as a function of stellar content (their stellar population). Closely related to CMD studies is the study of the density distribution of stars with different luminosities (the stellar luminosity function). In fact, luminosity functions are implicit in CMDs when the density of stars in each region of the CMD is considered, that is, the Hess diagram \citep{Hess24}. The 3D structure provided by the Hess diagram contains information about the stellar content of the system studied.

It was the development of stellar evolution theory in the 1950s \citep[e.g.][]{SS52} that allowed us to relate the observed structures in CMDs to the age and metallicity of the system, and explain the density of stars in the different areas of the Hess diagram according to the lifetime of stellar evolutionary phases (and hence the nuclear fuel in each phase). CMDs are divided into the main sequence (MS) region, with hydrogen-burning stars, and post-main sequence (PMS) stars. The stellar density of MS stars in a CMD depends on the stars formed at a given time with a given metallicity. The density of PMS stars in different areas depends on the lifetime of different evolutionary phases after MS turn-off. Stellar evolution theory allows us to transform current observable quantities into initial conditions and hence provides the frequency distribution of the properties of stars at birth. This frequency distribution, when expressed in probability terms, leads to the stellar birth rate, ${\cal B}(m_0,t,Z)$, which provides the probability that {\it a star} was born with a given initial mass, at a given time, and with a given metallicity.

By applying stellar evolution to the stellar luminosity function observed in the solar neighbourhood, \citet{Sal55} inferred the distribution of initial masses, the so-called initial mass function (IMF, $\phi(m_0)$). To quote \citet{Sal55}, `This luminosity function depends on three factors: (i) $\phi(m_0)$, the relative probability for the creation of stars near $m_0$ at a particular time; (ii) the rate of creation of stars as a function of time since the formation of our galaxy; and (iii) the evolution of stars of different masses.' There are various implications in this set of assumptions that merit detailed analysis. The first is that the luminosity function used in the work is not directly related to a stellar cluster, where a common physical origin would be postulated, but to a stellar ensemble where stars would have been formed under different environmental conditions. Implicitly, the concept of the stellar birth rate is extended to any stellar ensemble, independently of how the ensemble has been chosen. The second implication is that the stellar birth rate is decomposed into two different functions, the IMF and the star formation history (SFH, $\psi(t,Z)$). This is the most important assumption made in the modelling of stellar populations since it provides the definition of SFH and IMF, and our current understanding of galaxy evolution is based on such an assumption. Third, the IMF is a probability distribution; that is, the change from a (discrete) frequency distribution obtained from observations to a theoretically continuous probability distribution \citep[][among others]{Sal55,Math59,Sca86}.
The direct implication of the IMF definition is that, to quote \citet{Sca86}, there is `no means of obtaining an empirical mass distribution which corresponds to a consistent definition of the IMF and which can be directly related to theories of star formation without introducing major assumptions.' I refer interested readers to \citet{Ceretal13a} for further implications that are usually not considered in the literature.

Following the historical developments, \citet{Hod89} defined the concept of a `population box', a 3D histogram in which the $x$--$y$ plane is defined by the age of the stars and their metallicity. The vertical axis denotes the number of stars in each $x$--$y$ bin, or the sum of the initial masses of the stars in the bin. It is related to the star formation history, $\psi(t,Z)$, and, ultimately, to the stellar birth rate. Fig.~\ref{fig:CMDcarma} shows the population box for a field in the Large Magellanic Cloud disc taken from \citet{Mesetal13}.

The population box of a system is the very definition of its stellar population; it comprises information relating to the different sets of stars formed at a given time with a given metallicity. The objective of any stellar population model is to obtain such population boxes. The model is obtained at a global level, restricted to subregions of the system, or restricted to particular $x$--$y$ bins of the population box.

A stellar ensemble with no resolved components does not provide a CMD, but the sum of all the stars in the CMD. However, we can still infer the stellar population of the system by using just this information if we can characterize the possible stellar luminosity functions that sum results in the integrated luminosity of the system. The problem of inferring the stellar content from the integrated light of external galaxies was formalized by \citet{Whi35}. The idea is to take advantage of different photometric or particular spectral characteristics of the stars in the different regions of a CMD and combine them in such a way that we can reproduce the observations. The only requirement is to establish, on a physical basis, the frequency distribution of the spectral types and absolute magnitudes of the stars in the CDM. Studies by Sandage, Schwarzschild and Salpeter provide the probability distribution (instead of a frequency distribution) needed for proper development of Whipple's ideas, and in the late 1960s and the 1970s, mainly resulting from the work of \citet{Tins68,Tins72, TG76,Tins80}, a framework for evolutionary population synthesis and chemical evolution models was established.

In the 1980s, there were several significant developments related to stellar populations. The first was the definition of the single-age single-metallicity stellar population (SSP) by \citet{Ren81} \citep[see also][]{RB83}. An SSP is the stellar birth rate when the star formation history is a Dirac delta distribution \citep[see][for an extensive justification of such an approximation]{Buzz89}. In addition, an SSP is each one of the possible points in the age--metallicity plane of the population box. The second development resulted from work by \citet{CB91} and \citet{BC93} using isochrone synthesis. The density of PMS stars in an SSP depends mainly on the lifetime of each PMS phase, which is linked to the amount of fuel in the phase. In addition, all PMS stars in an SSP have a mass similar to that of the MS turn-off, $m_\mathrm{TO}$. Isochrone synthesis assumes that the similar recipes used to interpolate tracks in main sequence stars (homology relations in polytropic models) are valid for PMS stars as long as interpolations are between {\it suitable} equivalent evolutionary points. Hence, each PMS phase at a given age is related to stars with initial mass $m_\mathrm{TO}+\Delta m$, and the density of stars in the phase is given by the integral of the IMF over the $\Delta m$ interval. I refer interested readers to \citet{MG01} for additional aspects of isochrone synthesis and the lifetime of PMS phases. The connection of the SSP concept to isochrones in terms of CMD diagrams and stellar luminosity functions implies that an SSP is the set of stars defined by an isochrone {\it including} the density of stars at each point of the isochrone when weighted by the IMF.

In the 1980s there were also several advances in the study of the so-called sampling effects in population synthesis, especially in relation to the number of PMS stars in a stellar population. \citet{BB77} investigated how evolutionary model results vary depending on the number of stars in a synthetic cluster. The authors reported that sampling effects originate from the way the IMF is sampled {\it and} large changes in the effective temperature of a star during the PMS evolution, when there are rapid evolutionary phases (situations in which small variations in $m_0$ produce wide variations in luminosity). Similar studies have been carried out by \citet{BM83,CBB88,GB93,GCBB95,SF97}. A common characteristic in these studies, besides the application to LMC clusters, is the identification of sampling effects with the occurrence by number of luminous PMS stars dominating the integrated light.

From a more global perspective, \citet{Buzz89} established a direct analytical formalism to evaluate sampling effects in clusters of different sizes by defining the effective number of stars, $N_\mathrm{eff}$, which contributes to the integrated luminosity\footnote{Monochromatic $N_\mathrm{eff}$ values for old stellar populations can be found at {\tt http://www.bo.astro.it/$\sim$eps/home.html} and {\tt http://www.iaa.es/$\sim$rosa/research/synthesis/HRES/ESPS-HRES.html} \citep{GDetal05} for young and intermediate SSPs.} (which varies with the age and wavelength considered). To do so, he assumed that the number of stars with a given mass follows a Poisson statistic; hence, dispersion of the total luminosity is the sum of the independent Poisson statistics for the number of stars with a given luminosity. Obviously, the main contribution to global dispersion of the given luminosity is caused by sparse but luminous stars. Independently, \citet{TS88}, using similar arguments about Poisson statistics, proposed the use of dispersion of the total flux of a galaxy image as a primary distance indicator (so-called surface brightness fluctuation, SBF), which is an {\it observational} quantity. In fact, $N_\mathrm{eff}$ and SBFs are related, as shown by \citet{Buzz93}. However, we must recall that theoretical SBFs are a measure of possible fluctuations around a mean value, a more general concept than that used by \citet{TS88}.

SBFs observations are one of the smoking guns of the stochasticity of stellar populations at work in nature, and the first case to take advantage of such stochasticity to draw inferences about the physical quantities of stellar systems. Another application of SBFs is the breaking of age--metallicity degeneracy in SSP results for old ages \citep{W94}; additional applications have been described by \citet{Buz05}.

Another aspect of SBFs is that they had been extensively studied and used in systems with {\it old} stellar populations. It would be surprising if we related stochasticity to IMF sampling: the integrated light in old stellar populations is dominated by low-mass stars, which are more numerous than high-mass ones. Hence, if stochasticity is related to the IMF alone, we would expect dispersion to fade out as a system evolves. In fact, this is an erroneous interpretation of sampling effects: as cited before, the number of stars in a given PMS phase is defined by the $\Delta m$ interval and the size of such an interval does not depend on the IMF.

To explain stochasticity in terms of Poisson statistics was the 
norm in the modelling of the physical scatter of stellar populations up to 2005 \citep[][among others]{LM00,Cetal01,CVGLMH02,GLB04}. The Poisson distribution, since it is discrete, is easily related to a natural number of stars and to interpretation of the IMF as a frequency distribution (as opposed to a probability density function). But it was modelling in the X-ray domain by \citet{GGS04} that established the key point of stochastic modelling by using the stellar luminosity function. The ideas of \citet{GGS04} were expanded to the optical domain by \citet{CLCL06}, who thereby established a unified formalism in the modelling of stellar populations that can be applied from CMDs to the integrated light of galaxies. This work showed that the Poisson statistic is invalid in so far as the total number of stars in a system are correlated to each other by the ${\cal B}(m_0,t,Z)$ probability distribution and introduced covariance terms in the computation of the scatter by synthesis models. Such covariance terms are especially relevant in SBF computation, as shown by \citet{CLJ08}. That study also showed that probability distributions, when expressed as frequency distributions, follow a multinomial distribution, which is the natural result of binning of the stellar luminosity function.

In a phenomenological approach, the scatter of synthesis models has been studied in Monte Carlo simulations, particularly since the late 1990s, although previous studies can be found in the literature \citep[e.g.][as well as work related to LMC clusters quoted earlier]{MHK91,CMH94}. There is a wide variety of such studies, ranging from simulations applied to specified targets or wavelength domains to general studies of Monte Carlo simulation results. Examples of specific domains include globular clusters \citep[old populations with ${\cal N} \sim 10^6$ stars by][among others]{Yoetal06,CB07}, $\gamma$-ray and optical emission from young clusters in our galaxy \citep{CLC00,Knoetal02,Vosetal09} and the study of SBF \citep{BCC98,RBCC05}. Examples of general studies include work by \citet{Broetal99,Brutuc,BC03,CLCL06,FL10,massclean,Pisetal11,SVL11,slug,bin}, among others. There are currently some public Monte Carlo synthesis codes 
for further research in the area, such as the {\it SLUG} package ({\tt {https://sites.google.com/site/runslug/}}) \citep{slug} and the {\it MASSCLEAN} packages ({\tt {http://www.physics.uc.edu/}\%{7Ebogdan/massclean.html}}) \citep{massclean,masscleancolors,masscleannage}.

A natural effect of Monte Carlo modelling is direct visualization of the range of scatter in model results. However, the relevant results in Monte Carlo sets are the {\it distribution of results} instead of their range (I discuss this point later). Whatever the case, these phenomenological studies have opened up several questions on the modelling of stellar populations and their applications. Some of these questions are as follows. (a) What are the limitations of traditional synthesis models, especially for low values of $\cal N$ \citep[][among others]{CVG03,CL04,SVL11}? (b) What is actually computed by traditional synthesis models, how are they linked to Monte Carlo modelling, and are they the limit for ${\cal N} \rightarrow \infty$ (e.g. use of terms such as {\it discrete population synthesis} vs. {\it continuous population synthesis} by \citealt{FL10}, or {\it discrete IMF} vs. {\it continuous IMF} models by \citealt{Pisetal11}) ? (c) How can Monte Carlo modelling be used to make inferences about stellar systems \citep{FLCW12,PHE12}? In this study I aim to solve some of these questions and provide additional uses of stellar population modelling. I begin with the modelling itself.

\section{The origin of stochasticity in stellar population modelling}
\label{sec:origin}

The stellar mass distribution in the fragmentation of a molecular cloud, even though it produces a discrete number of stars, has a continuous range of possible outcomes. Although we do not know the details of fragmentation and they vary from one cloud to another, we observe that the larger the stellar mass considered, the lower is the number of stars in the mass range near such a stellar mass. From the observed frequency distributions we make an abstraction to a continuous probability density distribution (statistical inference) such as the IMF or the stellar birth rate, which has proved to be a useful approach in characterizing this physical situation. By using such probability distributions, we construct the theory of stellar populations to obtain the possible luminosities of stellar ensembles (probabilistic description, the forward problem of {\it predicting results}) and produce a continuous family of probability distributions that vary according to certain parameters (mainly $\cal N$, $t$ and $Z$). Finally, we compare these predictions with particular observations aimed at {\it false } regions of the parameter space; that is, to obtain combinations of $\cal N$, $t$ and $Z$ that are {\it not compatible} with observations (hypothesis testing in the space of observable luminosities).

The previous paragraph summarizes the three different steps in which stellar populations are involved. These three steps require stochastic (or random) variables, but, although related, each step needs different assumptions and distributions that should not be mixed up. Note that in the mathematical sense, a random variable is one that does not have a single fixed value, but a set of possible values, each with an associated probability distribution.

\begin{enumerate}
\item {\it IMF inference.} This is a statistical problem: inference of an unknown underlying probability distribution from observational data with a discrete number of events. This aspect is beyond the scope of our review. However, IMF inferences require correct visualization of the frequency distribution of stellar masses to avoid erroneous results \citep{DAS86,MAU05}. Such a problem in the visualization of distributions is general for inferences independently of the frequency distribution (stellar masses, distribution of globular clusters in a galaxy, or a set of integrated luminosities from Monte Carlo simulations of a system with given physical parameters).

\item {\it Prediction of stellar population observables for given physical conditions}. This is a probabilistic problem for which the underlying probability distribution, IMF and/or ${\cal B}(m_0,t,Z)$ is obtained by hypothesis, since it defines the initial physical conditions. Such probability distributions are modified (evolved) to obtain a new set of probability distributions in the observational domain. The important point is that, in so far as we are interested in generic results, we must renounce particular details. We need a description that covers {\it all possible} stellar populations quantitatively, but not necessarily any particular one. This is reflected in the inputs used, which can only be modelled as probability distributions. Hence, we have two types of probability distributions: one defining the initial conditions and one defining observables.

Regarding the initial conditions, the stellar mass is a continuous distribution and hence it cannot be described as a frequency distribution but as a continuous probability distribution, that is, a continuous probability density function (PDF). Hence, the IMF does not provide `the number of stars with a given initial mass', but, after integration, the probability that a star had an initial mass within a given mass range. If the input distribution refers to the probability that {\it a star} has a given initial mass, the output distribution must refer to the probability that {\it a star} has a given luminosity (at a given age and metallicity).

For the luminosity of an ensemble of stars, we must consider {\it all possible combinations of the possible luminosities of the individual stars in the ensemble}. Again, this situation must be described by a PDF. In fact, the PDF of integrated luminosities is intrinsically related to the PDF that describes the luminosities of individual stars. A different question is how such a PDF of integrated luminosities is provided by synthesis models: these can produce a parametric description of the PDFs (traditional modelling), a set of luminosities from which the {\it shape} of the PDF can be recovered (Monte Carlo simulation), or an explicit computation of the PDF (self-convolution of the stellar PDF; see below).

\item {\it Inference about the physical conditions from observed luminosities}. This is a hypothesis-testing problem, which differs from statistical problems in the sense that it is not possible to define a universe of hypotheses. This implies that we are never sure that the best-fit solution is the actual solution. It is possible that a different set of hypotheses we had not identified might produce an even better fit. The only thing we can be sure of in hypothesis-testing problems is which solutions are incompatible with observed data. We can also evaluate the degree of compatibility of our hypothesis with the data, but with caution. In fact, the best $\chi^2$ obtained from comparison of models with observational data would be misleading; the formal solution of any $\chi^2$ fit is the whole $\chi^2$ distribution. \citet{tarantola} provide a general view of the problem, and \citet{FLCW12} (especially Section 7.4 and Fig. 16) have described this approach for stellar clusters.

Assuming the input distributions is correct and we aim to obtain only the evolutionary status of a system, we must still deal with the fact that the possible observables are distributed. An observed luminosity (or a spectrum) corresponds to a different evolutionary status and the distribution of possible luminosities (or spectra) is defined by the number of stars in our resolution element, $\cal N$. Since we have an accurate description of the distribution of possible luminosities as a PDF, the best situation corresponds to the case in which we can sample such a distribution of possible luminosities with a larger number of elements, $n_\mathrm{sam}$, with the restriction that the total number of stars, $N_\mathrm{tot}$, is fixed. As pointed out before, it is CMDs that have ${\cal N} = 1$ and $n_\mathrm{sam} = N_\mathrm{tot}$. We can understand now that the so-called sampling effects are related to $n_\mathrm{sam}$ and how we can take advantage of this by managing the trade-off between $n_\mathrm{sam}$ and $\cal N$. Note that the literature on sampling effects usually refers to situations in which $N_\mathrm{tot}$ has a low value; obviously, this implies that $n_\mathrm{sam}$ is low. However, that explanation loses the advantages that we can obtain by analysis of the scatter for the luminosities of stellar systems.

\end{enumerate}

In summary, {\it any time a probability distribution is needed, and such a probability distribution is reflected in observational properties, the description becomes probabilistic.} Stochasticity, involving descriptions in terms of probability distributions, is intrinsic to nature and is implicit in the modelling of stellar populations. It can be traced back to the number of available stars sampled for the IMF or the ${\cal B}(m_0,t,Z)$ distributions, but it is misleading to talk about IMF or stellar-birth-rate sampling effects, since these input distributions are only half of the story.

\subsection{From the stellar birth rate to the stellar luminosity function}

The aim of stellar population modelling is to obtain the evolution of observable properties for given initial conditions ${\cal{B}}(m_0,t,Z)$. The observable property may be luminosities in different bands, spectral energy distributions, chemical yields, enrichment, or any parameter that can be related to stellar evolution theory. Thus, we need to transform the initial probability distribution ${\cal{B}}(m_0,t,Z)$ in the PDF of the observable quantities, $\ell_i$, at a given time, $t_\mathrm{mod}$. In other words, we need to obtain the probability that in a stellar ensemble of age $t_\mathrm{mod}$, a randomly chosen star has given values of $\ell_1, \dots, \ell_n$. Let us call such a PDF $\varphi(\ell_1, \dots,\ell_n; t_\mathrm{mod})$, which is the theoretical stellar luminosity function. {\it The stellar luminosity function is the distribution needed to describe the stellar population of a system.}

Any stellar population models (from CMD to integrated luminosities or spectra), as well as their applications, have $\varphi(\ell_1, \dots,\ell_n;t_\mathrm{mod})$ as the underlying distribution. ${\cal{B}}(m_0,t,Z)$ and stellar evolution theory are the gateway to obtaining $\varphi(\ell_1, \dots,\ell_n; t_\mathrm{mod})$. However, in most stellar population studies, $\varphi(\ell_1, \dots,\ell_n;t_\mathrm{mod})$ is not computed explicitly; it is not even taken into consideration, or even mentioned. Neglecting the stellar luminosity function as the underlying distribution in stellar population models produces tortuous, and sometimes erroneous, interpretations in model results and inferences from observational data.

There are several justifications for not using $\varphi(\ell_1, \dots,\ell_n; t_\mathrm{mod})$; it is difficult to compute explicitly (see below) and it is difficult to work in an $n$-dimensional space in both the mathematical and physical senses. We can work directly with ${\cal{B}}(m_0,t,Z)$ and stellar evolution theory and isochrones, $\ell_i(m_0;\tau,Z)$, where $\tau$ is the time measured since the birth of the star, without explicit computation of $\varphi(\ell_1, \dots,\ell_n;t_\mathrm{mod})$. In the following paragraphs I present a simplified description of a simple luminosity $\varphi(\ell; t_\mathrm{mod})$. Such a description is enough for understanding most of the results and applications of stellar population models.

First, the physical conditions are defined by ${\cal{B}}(m_0,t,Z)$ but we must obtain the possible luminosities at a given $t_\mathrm{mod}$. When $t_\mathrm{mod}$ is greater than the lifetime of stars above a certain initial mass, the first component of $\varphi(\ell; t_\mathrm{mod})$ is a Dirac delta function at $\ell = 0$ that contains the probability that a randomly chosen star taken from ${\cal{B}}(m_0,t,Z)$ is a dead star at $t_\mathrm{mod}$. With regard to isochrones, dead stars are rarely in the tables in so far as only luminosities are included. However, they are included if the isochrone provides the cumulative ejection of chemical species from stars of different initial masses (as needed for chemical evolution models).

Second, stars in the MS are described smoothly by $\ell(m_0;\tau,Z)$, which are monotonic continuous functions. In this region there is a one-to-one relation between $\ell$ and $m_0$, so $\varphi(\ell; t_\mathrm{mod})$ resembles ${\cal{B}}(m_0,t,Z)$. For an SSP case with a power law IMF, $\varphi(\ell; t_\mathrm{mod})$ is another power law with a different exponent.

Third, discontinuities in $\ell(m_0;\tau,Z)$ lead to discontinuities or bumps in $\varphi(\ell; t_\mathrm{mod})$. Depending on luminosities before $\ell(m_{0-}; \tau, Z)$ and after $\ell(m_{0+}; \tau, Z)$, they lead to a gap in the distribution (when $\ell(m_{0-}; \tau, Z) < \ell(m_{0+}; \tau, Z)$), or an extra contribution to $\ell(m_{0+}; \tau, Z)$ (when $\ell(m_{0-}; \tau, Z) > \ell(m_{0+}; \tau, Z)$). Discontinuities are related to changes in stellar evolutionary phases as a function of the initial mass, and to intrinsic discontinuities of stellar evolution (e.g. stars on the horizontal branch).

Fourth, variations in the derivative of $\ell(m_0;\tau,Z)$ led to depressions and bumps in $\varphi(\ell; t_\mathrm{mod})$. 
Consider a mass range for which a small variation in $m_0$ leads to a large variation in $\ell$ (the so-called fast evolutionary phases). We must cover a large $\ell$ range with a small range of $m_0$ values; hence, there will be a lower probability of finding stars in the given luminosity range (and in such an evolutionary phase), and hence a depression in $\varphi(\ell; t_\mathrm{mod})$. The steeper the slope, the deeper is the depression. Conversely, when the slope of $\ell(m_0;\tau,Z)$ is flat, there is a large mass range sharing the same luminosity; hence, it is easier to find stars with the corresponding $\ell$ values, and the probability for such $\ell$ values increases. Both situations are present in PMS phases.

In summary: (a) $\varphi(\ell_1, \dots,\ell_n; t_\mathrm{mod})$ has three different regimes, a Dirac delta component at zero luminosity because of dead stars, a low luminosity regime corresponding to the MS that resembles ${\cal{B}}(m_0,t,Z)$, and a high luminosity regime primarily defined by the PMS evolution that may overlap the MS regime; and (b) the lower the possible evolutionary phases in the PMS, the simpler is $\varphi(\ell_1, \dots,\ell_n; t_\mathrm{mod})$. In addition, the greater the possible evolutionary phases, the greater is the possibility of the incidence of bumps and depressions in the high-luminosity tail of the distribution.

\subsection{From resolved CMD to integrated luminosities and the dependence on $\cal N$}

\begin{figure}
\resizebox{\hsize}{!}{\includegraphics[angle=-90]{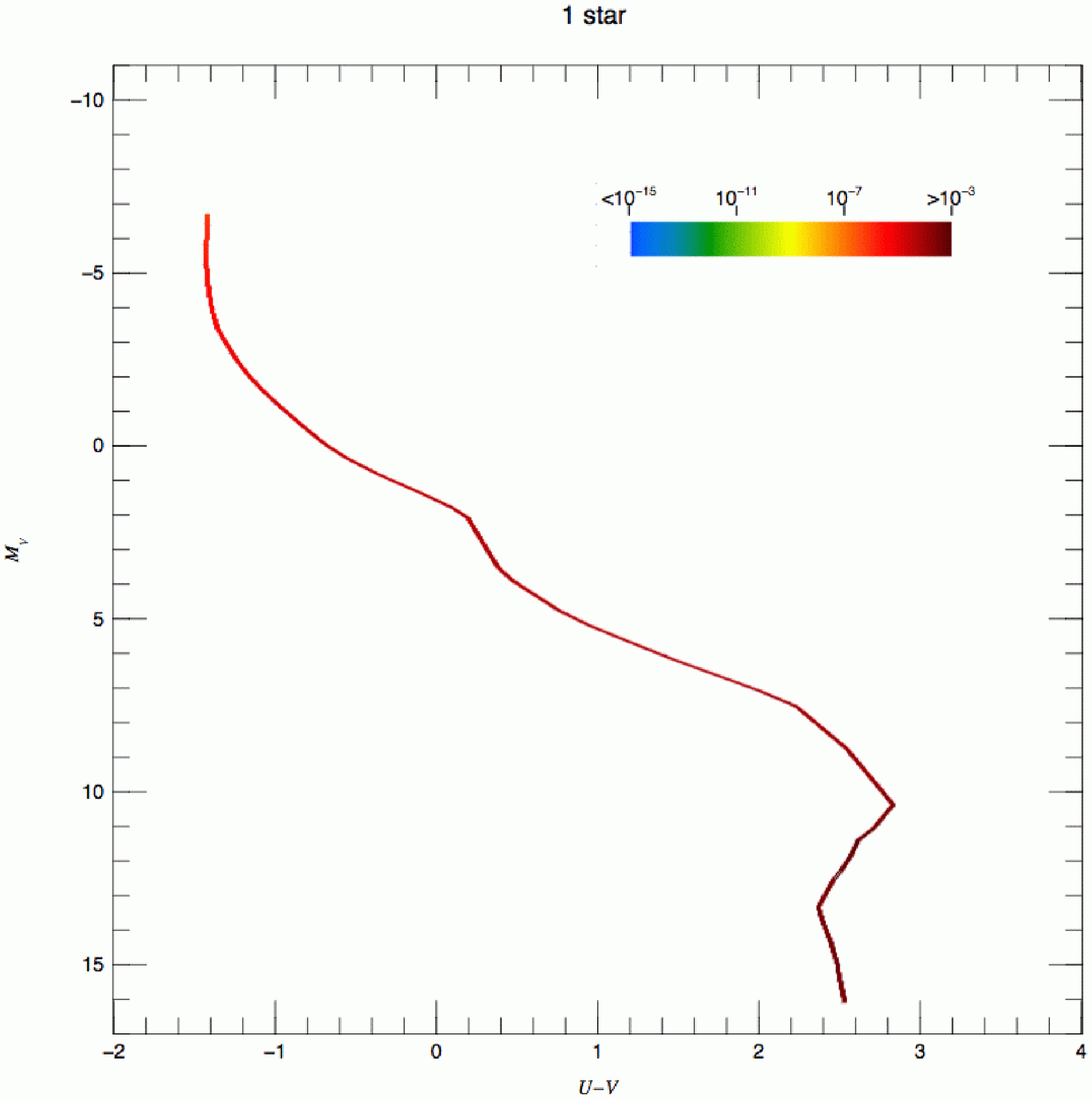}\includegraphics[angle=-90]{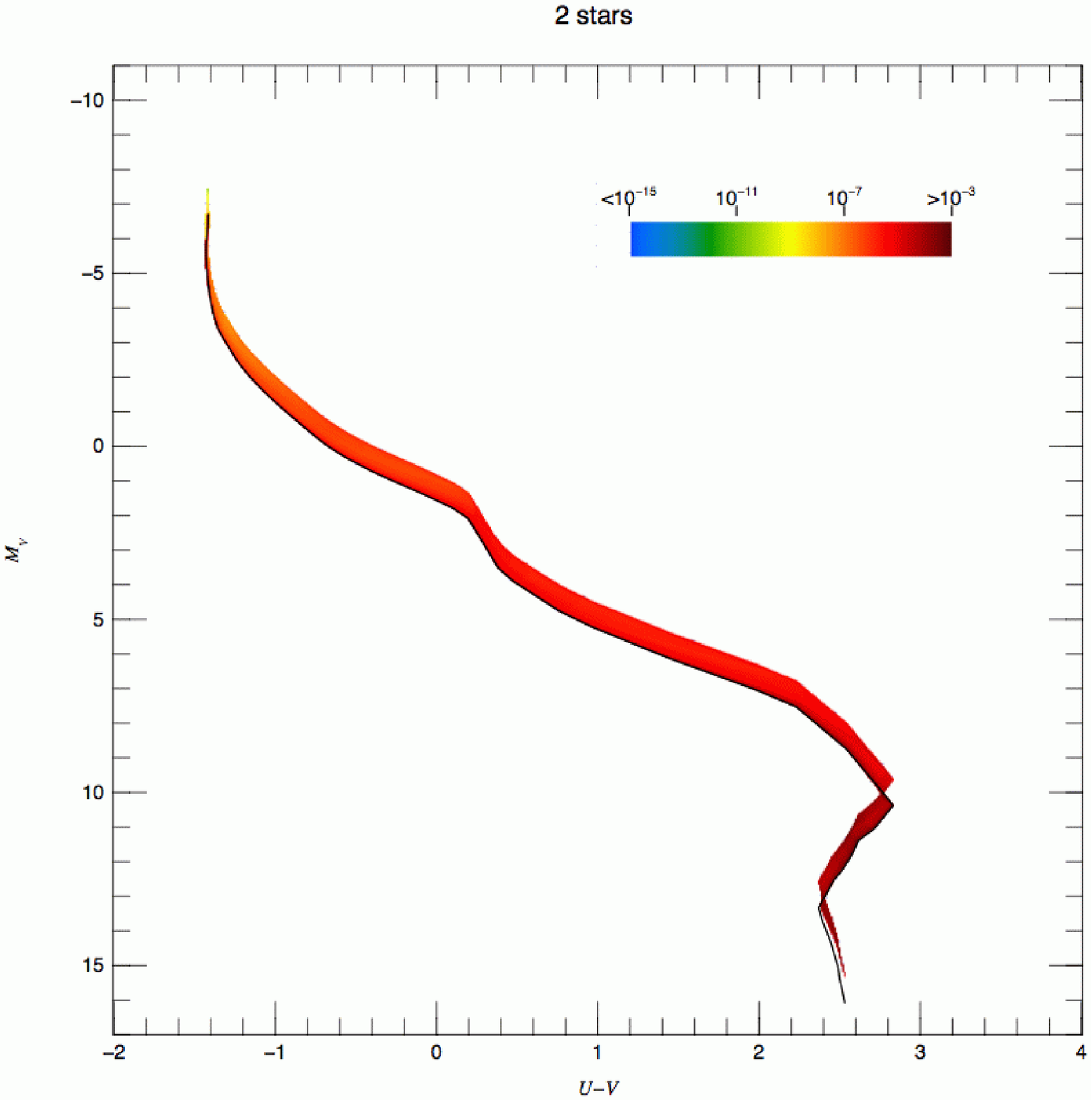}}\\
\resizebox{\hsize}{!}{\includegraphics[angle=-90]{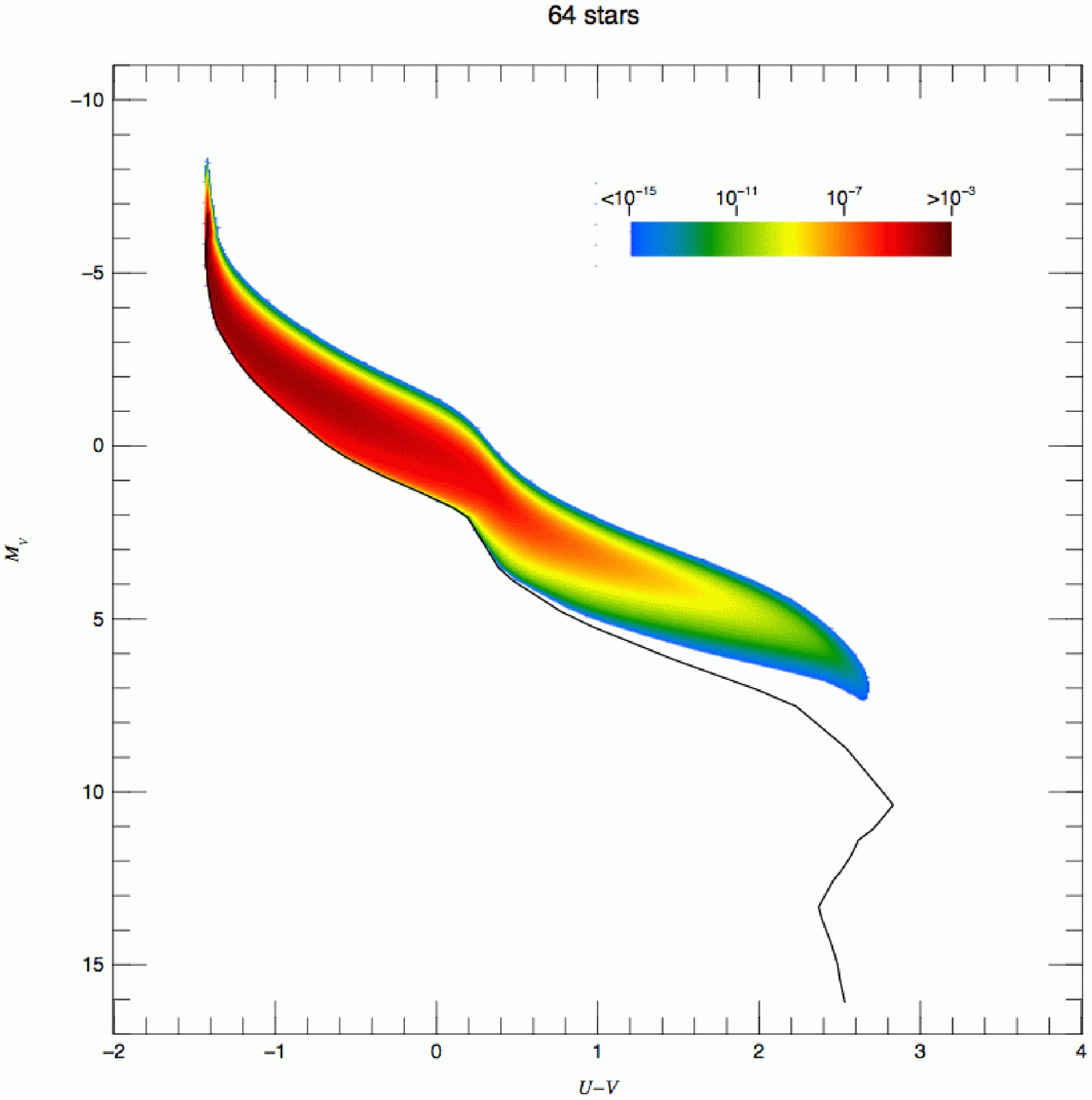}\includegraphics[angle=-90]{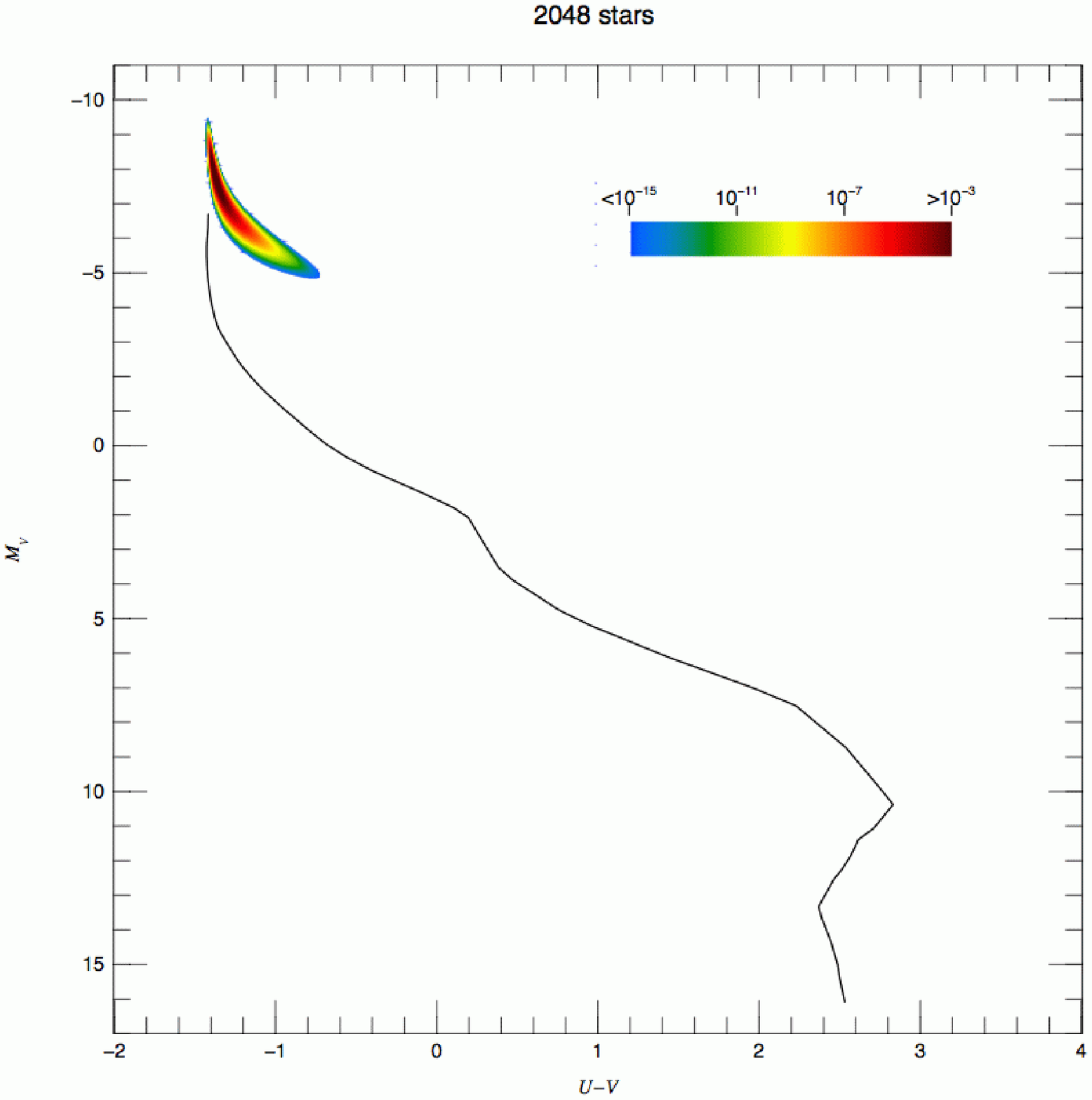}}
\caption[]{$U-V$ L versus $M_V$ Hess diagram for a zero-age SSP system with ${\cal N} = 1$ (top left), ${\cal N} = 2$ (top right), ${\cal N} = 64$ (bottom left) and ${\cal N} = 2048$ (bottom right) based on work by \citet{Maiz08}. Figure courtesy of J. Ma{\'{\i}}z Apell\'aniz.}
\label{fig:jma34}
\end{figure}

The previous section described the probability distribution that defines the probability that a randomly chosen star in a given system with defined ${\cal B}(m_0,t,Z)$ at time $t_\mathrm{mod}$ has given values of luminosity $\ell_1, \dots,\ell_n$. Let us assume that we now have a case in which we do not have all the stars resolved, but that we observe a stellar system for which stars have been grouped (randomly) in pairs. This would be a cluster in which single stars are superposed or blended. The probability of observing an integrated luminosity $L_{i,{\cal N}=2}$ from the sum for two stars is given by the probability that the first star has luminosity of $\ell_i$ multiplied by the probability that the second star has luminosity of $L_{i,{\cal N}=2} - \ell_i$ considering all possible $\ell_i$ values; that is, convolution of the stellar luminosity function with itself \citep{CLCL06}. Following the same argument, it is trivial to see that the PDF describing the integrated luminosity of a system with $\cal N$ stars, $\varphi_{\cal N}(L_1, \dots,L_n; t_\mathrm{mod})$, is the result of $\cal N$ self-convolutions of $\varphi(\ell_1, \dots,\ell_n; t_\mathrm{mod})$.

The situation in a $U-V$ versus \ $M_V$ Hess diagram is illustrated in Fig.~\ref{fig:jma34} for ${\cal N} = 1, 2, 64$ and 2048. The figure is from \citet{Maiz08}, who studied possible bias in IMF inferences because of crowding and used the convolution process already described. The figure uses logarithmic quantities; hence, it shows the {\it relative} scatter, which decreases as $\cal N$ increases. An important feature of the plot is that the case with a larger number of stars shows a banana-like structure caused by correlation between the $U$ and $V$ bands. In terms of individual stars, those that dominate the light in the $V$ band are not exactly the same as those that dominate the light in the $U$ band. However, there is partial correlation between the two types of stars resulting from ${\cal B}(m_0,t,Z)$. Such partial correlation is present when stars are grouped to obtain integrated luminosities. These figures also illustrate how CMD studies (${\cal N} = 1$) can be naturally linked to studies of integrated luminosities (${\cal N} > 2$) as long as we have enough observations ($n_{\mathrm{sam}}$) to sample the CMD diagram of integrated luminosities.

We now know that we can describe stellar populations as $\varphi_{\cal N}(L_1, \dots,L_n; t_\mathrm{mod})$ for all possible $\cal N$ values. The question now is how to characterize such a PDF. We can do this in several ways. (a) We can obtain PDFs via the convolution process, which is the most exact way. Unfortunately, this involves working in the $n$D space of observables and implementation of $n$-dimensional convolutions, which are not simple numerically. (b) We can bypass the preceding issues using a brute force methodology involving Monte Carlo simulations. This has the advantage (among others; see below) that the implicit correlations between the $\ell_i$ observables are naturally included and is thus a suitable phenomenological approach to the problem. The drawbacks are that the process is time-consuming and requires large amounts of storage and additional analysis for interpretation tailored to the design of the Monte Carlo simulations. This is discussed below. (c) We can obtain parameters of the PDFs as a function of $\cal N$. In fact, this is the procedure that has been performed in synthesis codes since their very early development: mean values and sometimes the variance of the PDFs expressed as SBF or $N_\mathrm{eft}$ are computed. The drawback is that a mean value and variance are not enough to establish probabilities (confidence intervals or percentiles) if we do not know the shape of the PDF.

\section{Stellar population modelling: the parametric approach and its limitations}
\label{sec:param}

\begin{figure}
\resizebox{\hsize}{!}{\includegraphics{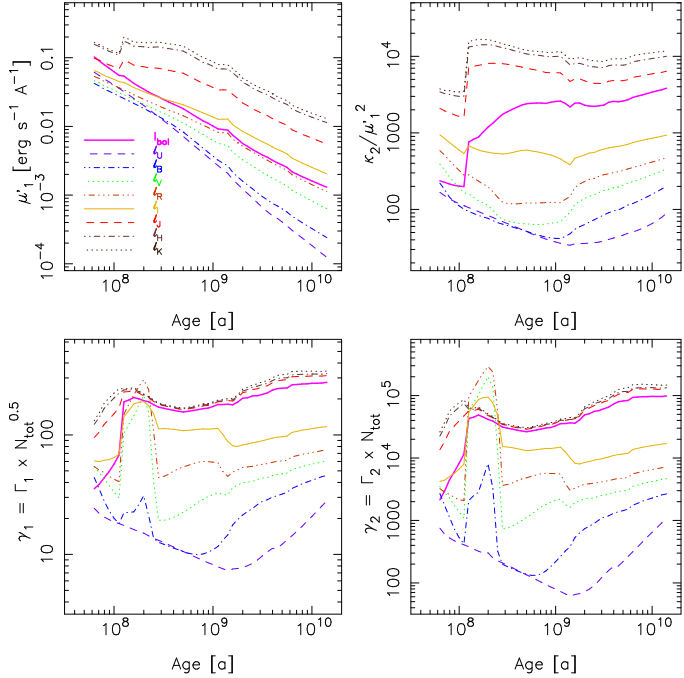}}
\caption[]{Main parameters of the luminosity function in several photometric bands. Figure from \citet{CLCL06}. In the notation used in this paper, $\kappa_2$ refers to $\mu_2$, $\Gamma_1$ refers to $\gamma_{1;{\cal N}}(L_i)$, and $\Gamma_2$ refers to $\gamma_{2;{\cal N}}(L_i)$.}
\label{fig:clcl06}
\end{figure}

\begin{figure}
\resizebox{\hsize}{!}{\includegraphics{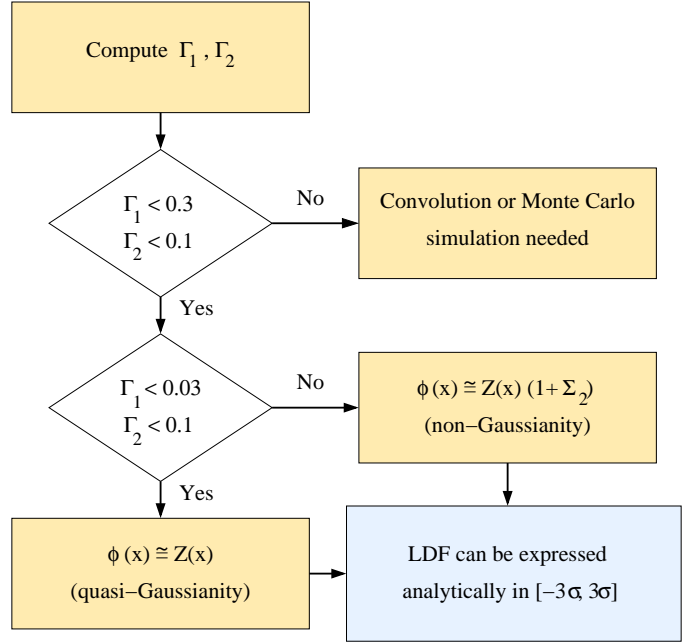}}
\caption[]{Characterization of a PDF based on Edgeworth's approximation to the second order and a Gaussianity tolerance interval of $\pm$10\%. Figure taken from \citet{CLCL06}, where $\Gamma_1$ refers to $\gamma_{1;{\cal N}}(L_i)$ and $\Gamma_2$ refers to $\gamma_{2;{\cal N}}(L_i)$.}
\label{fig:gammas}
\end{figure}
The parametric description of the stellar luminosity PDF is the one used by far the most often, although usually only the mean value of the distribution is computed. In fact, when we weight the stellar luminosity along an isochrone with an IMF in an SSP, it is such a mean value, $\mu_1'(\ell_i)$, that is obtained \citep[see][for more details]{CLCL06}.

We can also evaluate by how much the possible $\ell_i$ values differ from the mean value. For example, the variance $\mu_2(\ell_i)$, which is the average of the square of the distance to the mean (i.e. e integral of $(\ell_i - \mu_1'(\ell_i))^2 \, \varphi(\ell_i)$ over the possible $\ell_i$ values). In general, we can compute the difference between the mean and the possible $\ell_i$ elements of the distribution using any power, $(\ell_i - \mu_1'(\ell_i))^n$, and the resulting parameter is called the central $n$-moment, $\mu_n(\ell_i)$. We can also obtain the covariance for two different luminosities $\ell_i$ and $\ell_j$ by computing $(\ell_i - \mu_1'(\ell_i))^n(\ell_j - \mu_1'(\ell_j))^m$ integrated over $\varphi(\ell_i,\ell_j)$, where linear covariance coefficients are obtained for the case $n = m = 1$.

Parametric descriptions of PDFs usually use few parameters: the mean, variance, skewness, $\gamma_1(\ell_i) = \mu_3/\mu_2^{3/2}$, and kurtosis $\gamma_2(\ell_i) = \mu_4/\mu_2^2 -3$. Skewness is a measure of the asymmetry of the PDF. Kurtosis can be interpreted as a measure of how flat or peaked a distribution is (if we focus the comparison on the central part of the distribution) or how fat the tails of the distribution are (if we focus on the extremes) when compared to a Gaussian distribution. For reference, a Gaussian distribution has $\gamma_1 = \gamma_2 = 0$. Typical values of the four parameters and their evolution with time for an SSP case are shown in Fig.~\ref{fig:clcl06} taken from \citet{CLCL06}.

Large positive $\gamma_1$ values indicate that stellar luminosity PDFs in the SSP case are L-shaped, and large positive $\gamma_2$ values indicate that they have fat tails. In fact, we noted in the previous section that the stellar luminosity function is an L-shaped distribution composed of a power-law-like component resulting from MS stars and a fat tail at large luminosities because of PMS stars. However, $\gamma_1$ and $\gamma_2$ computation provides us with a {\it quantitative} characterization of the distribution shape without an explicit visualization.

The parameters that describe the distribution of integrated luminosities, $\varphi_{\cal N}(L_1, \dots,L_n; t_\mathrm{mod})$, are related to those of the stellar luminosity function by simple scale relations \citep{CLCL06}:

\begin{eqnarray}
\mu'_{1;{\cal N}}(L_i) & = & {\cal N} \times \mu_1'(\ell_i),\label{eq:mean}\\
\mu_{2;{\cal N}}(L_i) & = & {\cal N} \times \mu_2(\ell_i),\label{eq:variance}\\
\gamma_{1;{\cal N}}(L_i) & = & \frac{1}{\sqrt{\cal N}} \times \gamma_1(\ell_i), \label{eq:ske}\\
\gamma_{2;{\cal N}}(L_i) & = & \frac{1}{\cal N} \times \gamma_2(\ell_i). \label{eq:kur}
\end{eqnarray} 

We can then obtain additional scale relations for the effective number of stars at a given luminosity, $N_{\mathrm{eff};{\cal N}}(L_i)$ \citep{Buzz89}, the SBF, $\bar{L}$ \citep{TS88,Buzz93}, and the correlation coefficients between two different luminosities, $\rho(L_i,L_j)$:

\begin{eqnarray}
N_{\mathrm{eff};{\cal N}}(L_i) & = & {\cal N} \times N_\mathrm{eff}(\ell_i) = {\cal N} \times \frac{\mu'^2_1(\ell_i)}{\mu_2(\ell_i)},\label{eq:neff}\\
\bar{L}_{i} & = & = \frac{\mu_2(\ell_i)}{\mu'_1(\ell_i)} ~~\forall {\cal N}, \label{eq:sbf}\\
\rho(L_i,L_j) & = & \rho(\ell_i,\ell_j) = \frac{\mathrm{cov}(\ell_i,\ell_j)}{\sqrt{\mu_2(\ell_i) \mu_2(\ell_j)}} ~~~\forall {\cal N}. \label{eq:rho}
\end{eqnarray} 

Direct (and simple) computation of the parameters of the distribution provides several interesting results. First, SBFs are a measure of the scatter that is independent of ${\cal N}$ and can be applied to any situation (from stellar clusters to galaxies) in fitting techniques. Second, correlation coefficients are also invariant about ${\cal N}$ and can be included in any fitting technique. Third, since the inverse of $N_{\mathrm{eff};{\cal N}}(L_i)$ is the relative dispersion, when ${\cal N} \rightarrow \infty$ the relative dispersion goes to zero, although the absolute dispersion (square root of the variance, $\sigma$) goes to infinity. Fourth, when ${\cal N} \rightarrow \infty$, $\gamma_{1;{\cal N}}(L_i) $ and $\gamma_{2;{\cal N}}(L_i)$ goes to zero, and hence the shape of the distribution of integrated luminosities becomes a Gaussian-like distribution (actually an $n$-dimensional Gaussian distribution including the corresponding $\rho(L_i,L_j)$ coefficients). We can also obtain the range of $\gamma_{1;{\cal N}}(L_i)$ and $\gamma_{2;{\cal N}}(L_i)$ values for which the distribution can be approximated by a Gaussian or an expansion of Gaussian distributions (such an Edgeworth distribution) for a certain luminosity interval. As reference values, the shape of PDFs where $\gamma_{1;{\cal N}} < 0.3$ or $\gamma_{2;{\cal N}} < 0.1$ are well described with these four parameters by an Edgeworth distribution; when $\gamma_{1;{\cal N}} < 0.03$ and $\gamma_{2;{\cal N}} < 0.1$, the PDFs are well described by a Gaussian distribution with the corresponding mean and variance. A Monte Carlo simulation or a convolution process is needed in other situations. The possible situations are shown in Fig.~\ref{fig:gammas}, taken from \citet{CLCL06}.

Another possibility that covers situations for higher $\gamma_{1;{\cal N}}$ and $\gamma_{2;{\cal N}}$ values quoted here (i.e. asymmetric PDFs) is to approximate $\varphi(\ell_i)$ by gamma distributions, as done by \citet{MA09}. This approach can be used in a wide variety of situations as long as the PDF has no bumps or the bumps are smooth enough and an accurate description of the tails of the distribution is not required.

\subsection{The mean and variance obtained using standard models}

We have seen that using Eq.~ (\ref{eq:mean}), the mean value can be expressed for any possible $\cal N$ value or for any quantity related to $\cal N$. Most population synthesis codes use the mass of gas transformed into stars or the star formation rate (also expressed as the amount of mass transformed into stars over a time interval) instead of referring to the number of stars. Hence, the typical unit of luminosity is [erg s$^{-1}$ M$_\odot^{-1}$] or something similar. However, here I argue that the computed value actually refers to the mean value of $\varphi_\ell(\ell)$, so the units of the luminosity obtained by the codes should be [erg s$^{-1}$] and refer to individual stars.

In fact, the difference is in the interpretation and algebraic manipulation of ${\cal B}(m_0,t,Z)$ and the IMF. The usual argument has two distinct steps. (1) The integrated luminosity and total stellar mass in a system are the sum of the luminosities and masses of all the individual stars in the system. Thus, the ratio of luminosity to mass produces the mass--luminosity relation for the system. (2) ${\cal B}(m_0,t,Z)$ (or the IMF) provides the {\it actual} mass of the individual stars in the system, and since the shape of such functions is independent of the number of stars in the system, the previous mass--luminosity relations are valid for any ensemble of stars with similar ${\cal B}(m_0,t,Z)$ functional form. The first step is always true as long as we know the masses and evolutionary status of all the stars in the system (actually it is the way that each {\it individual} Monte Carlo simulation obtains observables). The second step is false: {\it we do not know the individual stars in the system}. We can describe the set using a probability distribution, and hence we can describe the integrated luminosity of {\it all} possible combinations of a sample of ${\cal N}$ individual stars.

It is trivial to see that the mass normalisation constant used in most synthesis codes is actually the mean stellar mass $\mu'_1(m_0)$ obtained using the IMF as a PDF.\footnote{ For reference, a Salpeter IMF in the mass range 0.08--120~M$_\odot$ has $\mu'_1(m_0) = 0.28$ M$_\odot$, $\mu_2(m_0) = 1.44$, $\gamma_1(m_0) = 1691.47$, and $\gamma_2(m_0) = 2556.66$. This implies that the distribution of the total mass becomes Gaussian-like for ${\cal N} \sim 3\times 10^7$ stars when the average total mass is approximately $9\times 10^6$ M$_\odot$.} Equivalently, total masses or star formation rates obtained using inferences from synthesis models are actually ${\cal N} \times \mu'_1(m_0)$ and ${\cal N} \times \mu'_1(m_0) \times t^{-1}$. Usually, this difference has no implication; however, it is different to say that a galaxy has a formation rate of, say, 0.1 stars per year (it forms, on average, a star every 10 years whatever its mass) than 0.1 M$_\odot$ per year (does this mean that, on average, $10^3$ years are needed to form a 100-M $_\odot$ star without forming any other star?).

Different renormalisations are performed on a physical basis, depending on the system we are interested in. Low-mass stars in young starbursts make almost no contribution to the UV integrated luminosity, so we can exclude low-mass stars from the modelling. Massive stars are not present in old systems; hence, we do not need to include massive stars in these SSP models. The use of different normalisations can be solved easily using a renormalisation process. Hence, we can compare the mean values obtained from two synthesis codes that use different constraints. However, we must be aware that 
underlying any such renormalisation there are different constraints on ${\cal B}(m_0,t,Z)$, and hence there are changes in the shape of the possible $\varphi(\ell_1, \dots, \ell_n; t_\mathrm{mod})$, such as the absence or presence of a Dirac delta function at $\ell_i = 0$. This affects the possible values of the mean, variance (SBF or $N_\mathrm{eft}$), skewness, and kurtosis used to describe $\varphi_{\cal N}(L_1, \dots,L_n; t_\mathrm{mod})$.

\section{Stellar populations using Monte Carlo modelling}
\label{sec:montecarlo}

\begin{figure}
\resizebox{\hsize}{!}{\includegraphics[angle=-90]{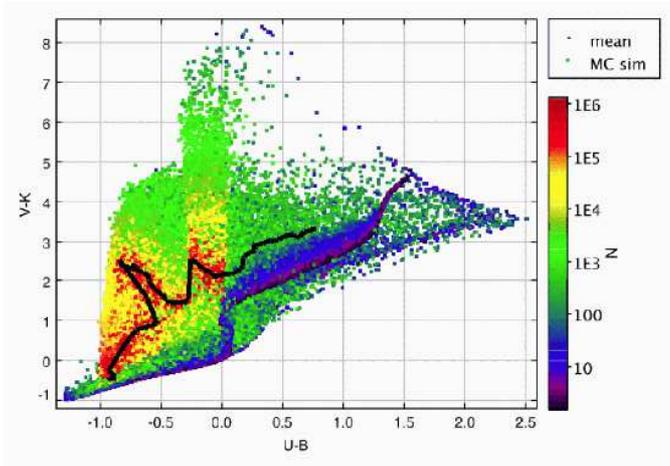}}
\caption[]{Results of $10^7$ Monte Carlo simulations of SSP models in the age range from 40 Myr to 2.8 Gyr, with $\cal N$ in the range between 2 and $10^6$ stars. The black line is the mean value of the colours obtained by a synthesis model in the age range from 40 Myr to 10 Gyr.}
\label{fig:mon1}
\end{figure}

\begin{figure}
\resizebox{\hsize}{!}{\includegraphics[width=0.15\hsize, angle=-90]{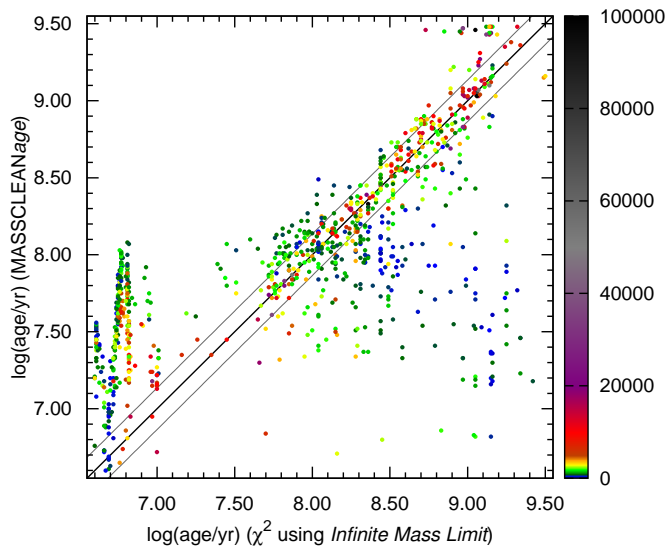}}
\caption[]{MASSCLEAN age \citep{masscleannage} results versus a $\chi^2$ minimisation fit to the mean value obtained from SSPs for age inferences of LMC clusters. The dots are colour-coded to show the mass of the cluster. Figure from \citet{PHE12}, who call the mean obtained by parametric models the infinite mass limitL.}
\label{fig:massclean}
\end{figure}

Monte Carlo simulations have the advantage (and the danger) that they are easy to do and additional distributions and constraints can be included in the modelling. Fig.~\ref{fig:mon1} shows the results of $10^7$ Monte Carlo simulations of SSP models with ages following a power-law distribution between 40 Myr and 2.8 Gyr, and $\cal N$ distributed following a (discrete) power-law distribution in the range between 2 and $10^6$ stars. The figure also shows the standard modelling result using the mean value of the corresponding observable in the age range from 40 Myr to 10 Gyr.

The first impression obtained from the plot is psychologically depressing. We had developed population synthesis codes to draw inferences from observational data, and the Monte Carlo simulations show so large a scatter that we wonder if our inferences are correct. Fig.~\ref{fig:massclean}, from \citet{PHE12}, compares the age inferences for LMC clusters using Monte Carlo simulations to sample the PDF of integrated luminosities and traditional $\chi^2$ fitting to the mean value of the PDF (traditional synthesis model results). The figure shows a systematic discrepancy at young ages. Such an effect was also found by \citet{FLCW12} for M83 stellar clusters and by \citet{FL10,SVL11} (among others) when trying to recover the inputs of the Monte Carlo simulations using $\chi^2$ fitting to the mean value obtained by parametric models. This requires stepwise interpretation. First, we must understand Monte Carlo simulation results. Second, we must figure out how to use Monte Carlo simulations to make inferences about observational data (hypothesis testing).

\subsection{Understanding Monte Carlo simulations: the revenge of the stellar luminosity function}

The usual first step is to analyse (qualitatively) the results for Monte Carlo simulations that implicitly represent how $\varphi_{\cal N}(L_1, \dots,L_n; t_\mathrm{mod})$ varies with $\cal N$ \citep{Brutuc,CVG03,massclean,masscleancolors,masscleannage,Pisetal11,FL10,SVL11,slug,bin}. In general, researchers have found that at very low $\cal N$ values, most simulations are situated in a region away from the mean obtained by parametric models; at intermediate $\cal N$ values, the distribution of integrated luminosities or colours are (sometimes) bimodal ; at large $\cal N$ values, the distributions become Gaussian.

\begin{figure}
\resizebox{\hsize}{!}{\includegraphics{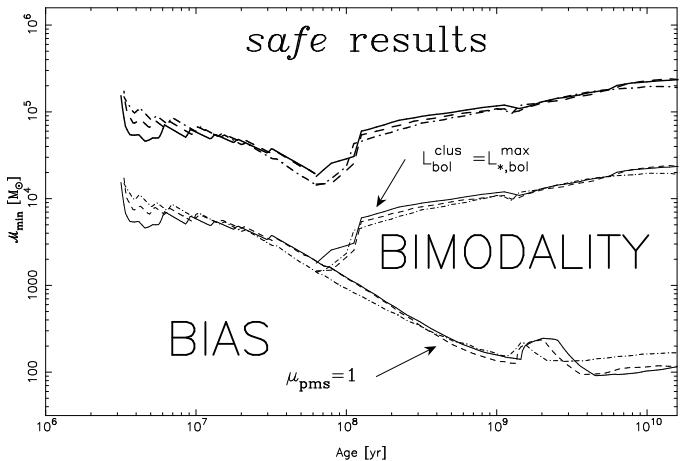}}
\caption[]{${\cal N} \times \mu_1'(m_0) = {\cal M}$ ranges comparing the mean and mode of the distributions of integrated luminosities for different situations. The plots were computed for three metallicities: $Z=0.019$ (solid line), $Z=0.008$ (dashed line) and $Z=0.0004$ (dotted dashed line). The upper line defines the ${\cal M}$ value for which there are at least 10 PMS stars ($t < 10^8$ yr) or where $L_\mathrm{bol}^\mathrm{clus} > 10 \times L_{*,\mathrm{bol} }^{\mathrm{max}}$ ($t > 10^8$). Figure from \citet{Cer03}.}
\label{fig:MonlocationC03}
\end{figure}

The different situations are shown quantitatively in Fig.~\ref{fig:MonlocationC03}, taken from \citet{Cer03}, which can be fully understood if we take advantage of the Monte Carlo simulations and the parametric description of synthesis models \citep[after][]{GGS04,CLCL06}. For ages greater than 0.1 Gyr, the figure compares when the mean bolometric luminosity of a cluster with $\cal N$ stars, $L_\mathrm{bol}^{\mathrm{clus}} ={\cal N}Ê\times \mu_1'(\ell_\mathrm{bol}; t_\mathrm{mod})$, equals the maximum value of $\varphi(\ell_\mathrm{bol}; t_\mathrm{mod})$, $L_{*,\mathrm{bol}}^{\mathrm{max}}$. This can be interpreted as the situation in which the light of a cluster is dominated by a single star, called the `lowest luminosity limit' by \citet{CL04}. In addition, the $\cal N$ value needed to have, on average, at least 1 PMS star is also computed; it can be obtained using a binomial distribution over the IMF for stars below and above $m_\mathrm{TO}$. The corresponding $\cal N$ values are expressed as the mean total mass of the cluster, $\cal M$.

\begin{figure}
\resizebox{\hsize}{!}{\includegraphics{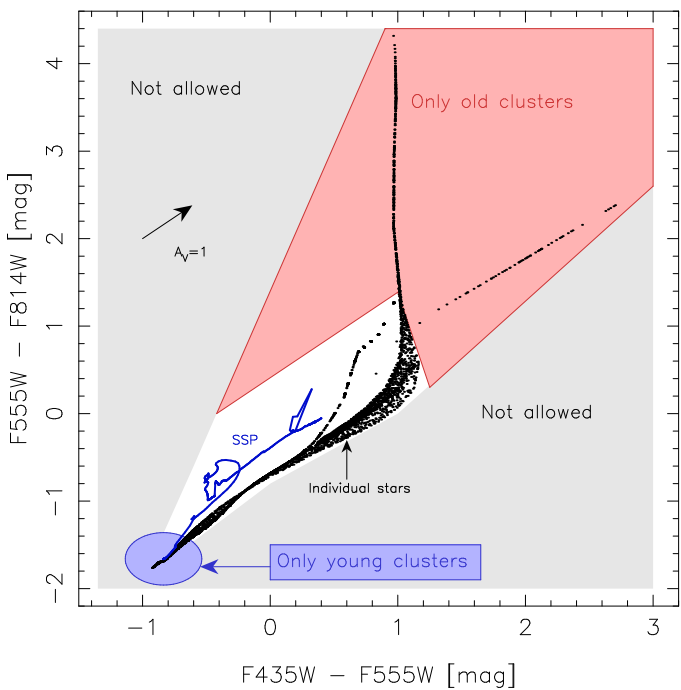}}
\caption[]{Graphical representation of the boundary conditions. The blue track shows the mean value obtained by SSP models (from the bottom-left corner of the figure to the red colour as a function of increasing age). The red area defines the region within which only old ($\ge 10^8$ yr) clusters can reside; similarly, the blue area is the region for young ($\le 5$ Myr-old) clusters. The black dots show the positions of individual stars. The grey shaded areas cannot host any clusters since there is no possible combination of single stars that can produce such cluster colours. The extinction vector for $A_V$ = 1 mag is shown for reference. Figure from \citet{BdGC08}.}
\label{fig:Monlocation}
\end{figure}

From the description of $\varphi(\ell;t_\mathrm{mod})$ we know that the distribution is L-shaped with a modal value (maximum of the distribution) in the MS region and a mean located somewhere between the MS and PMS regions. The mean number of PMS will be less than one at low $\cal N$ values. This means that most simulations are low-luminosity clusters composed of only MS stars, and a few simulations are high-luminosity clusters dominated by PMS stars. Here, the mode of $\varphi_{\cal N}(L ;t_\mathrm{mod})$ is defined by MS stars and is biased with respect to the mean of $\varphi_{\cal N}(L ;t_\mathrm{mod})$. We can also be sure that the distribution $\varphi_{\cal N}(L ;t_\mathrm{mod})$ is not Gaussian-like if its mean is lower than the maximum of $\varphi(\ell ;t_\mathrm{mod})$, since the shape of the distribution is sensitive to how many luminous stars are present in each simulation (e.g. 0, 1, 2, \dots) and this leads to bumps (and possibly multi-modality) in $\varphi_{\cal N}(L ;t_\mathrm{mod})$. Finally, when $\cal N$ is large, $\varphi_{\cal N}(L ;t_\mathrm{mod})$ becomes Gaussian-like and the mean and mode coincide. \citet{Cer03,CL04} established the maximum of $\varphi(\ell ;t_\mathrm{mod})$ 10 times to reach this safe result. Curiously, the value of $\cal N$ that can be inferred from Fig.~\ref{fig:MonlocationC03} can be used in combination with $\gamma_1$ and $\gamma_2$ values from Fig.~\ref{fig:clcl06}. The resulting $\gamma_{1:{\cal N}}$ and $\gamma_{2;{\cal N}}$ values are approximately 0.1, the values required for Gaussian-like distributions obtained by the parametric analysis of the stellar luminosity distribution function.

This exercise shows that interpretation of Monte Carlo simulations can be established {\it quantitatively} when the parametric description is also taken into account. Surprisingly, however, except for Monte Carlo simulations used to obtain SBFs \citep{Broetal99,RBCC05,GLB04}, most studies do not obtain the mean or compare it with the result obtained by parametric modelling. Usually, Monte Carlo studies only verify that for large $\cal N$ values, using relative values, the results coincide with standard models. In addition, hardly any Monte Carlo studies make explicit reference to the stellar luminosity function, nor do they compute simulations for the extreme case of ${\cal N} = 1$. Thus, psychological bias is implicit for interpretation of ${\cal B}(m_0,t,Z)$ referring only to stellar {\it clusters} and synthesis models referring only to {\it integrated} properties.

Another application of the stellar luminosity function is definition of the location of simulations/observations in colour--colour diagrams (actually in any diagnostic diagram using indices that do not depend on $\cal N$). Any integrated colour is a combination of the contributions of different stars; hence, individual stars define an envelope of possible colours of simulations/observations. The situation is illustrated in Fig.~\ref{fig:Monlocation} taken from \citet{BdGC08}, where the possible range of colours of individual stars and the mean of parametric synthesis models are compared.

The final application of parametric descriptions of the integrated luminosity function to Monte Carlo simulations is a back-of-the-envelope estimation of how many Monte Carlo simulations are needed for reliable results. This number depends on the simulation objective, but a minimal requirement is that, for a fixed $\cal N$, the mean value obtained from the simulations (the sample mean $\left< \tilde{L} \right>$) must be consistent to within an error of $\epsilon$ of the mean obtained by the parametric modelling (the population mean). Statistics textbooks show that, independent of the shape of the distribution, the sampling mean is distributed according to a sampling distribution with variance equal to the variance of the population distribution variance divided by the sample size:

\begin{equation}
\sigma^2(\left< \tilde{L} \right>) = {\frac{\mu_{2;{\cal N}}(L)}{n_\mathrm{sam}}} = \frac{\cal N}{n_\mathrm{sam}} \, \mu_{2}(\ell).
\end{equation}

Hence, expressed in relative terms, we can require that

\begin{equation}
\frac{\sigma(\left< \tilde{L} \right>)}{\mu_1'(L)} = \frac{1}{\sqrt{n_\mathrm{sam} {\cal N}}} \frac{\sqrt{\mu_{2}(\ell)}}{\mu_1'(\ell)} < \epsilon.
\end{equation}
We can impose a similar requirement for the variance obtained from the distribution, which results in

\begin{equation}
\frac{\sigma(\left< \tilde{\sigma}^2(L) \right>)}{\mu_{2;{\cal N}}(L)} \approx \frac{1}{\sqrt{n_\mathrm{sam} {\cal N}}} \sqrt{\frac{1}{{\cal N}} \gamma_2 +2} < \epsilon.
\label{eq:samvarvar}
\end{equation}
An interesting result is that, for relative dispersion, the relevant parameter is the total number of stars $N_\mathrm{tot}$ used in the overall simulation set. Hence, the number of simulations needed to sample the distribution of integrated luminosities decreases when $\cal N$ increases. However, for absolutes values, the ratio of ${\cal N}$ to $n_\mathrm{sam}$ is the relevant quantity.

\subsection{What we can learn from Monte Carlo simulations? }

The principal advantage of Monte Carlo simulations is that they allow us to include constraints that are difficult to manage with a parametric description. Examples are constraints in the inputs (e.g. considering only simulations with a given number of stars in a given (observed) mass range, \citealt{Knoetal02}) or constraints in the outputs (e.g. considering only simulations that verify certain observational constraints in luminosities, \citealt{Luretal03}). The issue here is to compute Monte Carlo simulations and only consider those that are consistent with the desired constraints. Note that once a constraint is included, the process requires transformation of the associated probability distributions and hence a change in the parameters of the distribution, and some constraints cannot be expressed analytically as a function of input parameters. Another application that can easily be performed with Monte Carlo simulations is testing of hypotheses by comparing the distribution of the simulations and observations; examples have been described by \citet{FdSK11,bin}. Hence, this approach is ideal when fine-tuned to observational (or theoretical) constraints.

However, at the beginning of this section I stated that one of the dangers of Monte Carlo simulations is that they are easy to do. They are so easy that we can include additional distributions, such as the distribution of numbers of stars in clusters, the total mass distribution, an age distributions, without a fine-tuned specific purpose. Here, the danger is that we must understand the questions that we are addressing and how such additional distributions affect the possible inferences. For instance, \citet{Yoetal06} used Monte Carlo simulations with the typical number of stars in a globular cluster to explain bimodal distributions in globular clusters. They argued that bimodality is the result of a non-linear relation between metallicity and colour transformation. The transformation undoubtedly has an effect on the bimodality; however, the number of stars used (actually the typical number of stars that globular clusters have) is responsible for the bimodal distributions. Monte Carlo simulations using large $\cal N$ values must converge to a Gaussian distribution.

When Monte Carlo simulations are used to obtain parameters for stellar clusters, the usual approach is to apply large grids covering the parameter space for $\cal N$ and age. However, the situation is not as simple as expected. First, a typical situation is to use a distribution of total masses. Hence, $\cal N$ is described by an unknown distribution. Even worse, $\cal M$ fluctuates since the simulations must include the constraint that $\cal N$ is an integer. Therefore, the mean values of such simulations diverge from the mean values obtained in the parametric description noted before, since they include additional distributions. Second, the inferences depend on the input distributions. For instance, \citet{PHE12} produced a grid assuming a flat distribution for total mass and a flat distribution in $\log t$. They compared observational colours with the Monte Carlo grid results and obtained a distribution of the parameters (age and mass) of the Monte Carlo simulations compatible with the observations. When expressed in probabilistic terms, the grid of Monte Carlo simulations represents the probability that a cluster has a given luminosity set for given age and mass, ${\cal P}(L | t, {\cal M})$, and comparison of observational data with such a set represents the probability that a cluster has a given age and mass for a given luminosity set ${\cal P}(t, {\cal M} | L)$. The method seems to be correct, but it is a typical fallacy of conditional probabilities: it would be correct only if the resulting distribution of ages were flat in $\log t$ and if the distribution of masses were flat. The set of simulations has a prior hypothesis about the distribution of masses and ages, and the results are valid only so far as the prior is realistic. In fact, \citet{PHE12} obtained a distribution of ages and total masses that differs from the input distributions used in the Monte Carlo simulation set.

The situation was also illustrated by \citet{FLCW12}, who computed a Monte Carlo set with a flat distribution in both $\log t$ and $\log {\cal M}$ as priors. The authors were aware of the Bayes theorem, which connects conditional probabilities:

\begin{equation}
{\cal P}(t, {\cal M} | L) = \frac{{\cal P}(L | t, {\cal M}) \, {\cal P}({\cal M})}{{\cal P}(L)}.
 \label{eq:bayes}
\end{equation}

They obtained age and mass distributions for observed clusters using a similar methodology to that of \citet{PHE12}. In fact, they found that the distribution of total masses follows a power law with an exponent $-2$; hence, the distribution used as a prior is false. However, they claimed that the real distribution of total masses is the one obtained, similar to \citet{PHE12}. Unfortunately, the authors were unaware that such a claim is valid only as long as a cross-validation is performed. This should involve repetition of the Monte Carlo simulations with a total mass distribution following an exponent of $-2$ and verification that the resulting distribution is compatible with such a prior \citep{tarantola}.

Apart from problems in using the Bayes theorem to make false hypotheses, both studies can be considered as major milestones in the inference of stellar populations using Monte Carlo modelling. Their age and total mass inferences are more realistic since they consider intrinsic stochasticity in the modelling, which is undoubtedly better than not considering stochasticity at all. The studies lack only the final step of cross-validation to obtain robust results.

Monte Carlo simulations provide two important lessons in the modelling of stellar populations that also occur in the Gaussian regime and can apply to systems of any size. The first is the problem of priors and cross-validation (i.e. an iteration of the results). The second and more important lesson is that there are no unique solutions; the best solution is actually the distribution of possible solutions. In fact, we can take advantage of the distribution of possible solutions to obtain further results \citep{FLCW12}.

\section{Implications of probabilistic modelling}
\label{sec:implications}

We have seen the implications of probabilistic modelling for a low-$\cal N$ regime when Monte Carlo simulations (or covolution of the stellar luminosity function) are required. However, we have shown that some characteristics of stochasticity are present independently of $\cal N$ (as in SBFs and partial correlations). In addition, we have seen that we can combine different probability distributions to describe new situations, as for integrated luminosities at a given $\cal N$ or integrated luminosities of clusters that follow a $\cal N$ or age distribution. Let us explore the implications of such results.

\subsection{Metrics of fitting}
\label{sec:metrics}

\begin{figure}
\resizebox{\hsize}{!}{\includegraphics[angle=-90]{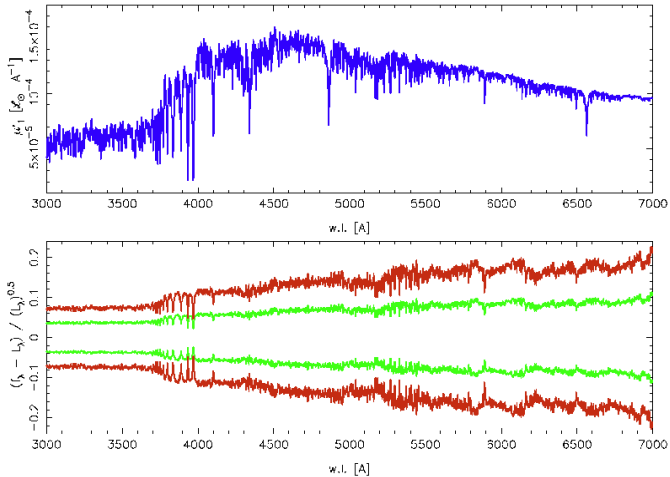}}
\caption[]{Top: Integrated mean spectra of a 1-Gyr-old cluster. Bottom: 1 and 2$\sigma$ confidence intervals for the mean-averaged dispersion $(L_\lambda - \mu_{1;N}'(L) / \sqrt(\mu_{1;N}'(L))$. Figure from \citet{CL09}.}
\label{fig:metrics}
\end{figure}

\begin{figure}
\resizebox{\hsize}{!}{\includegraphics{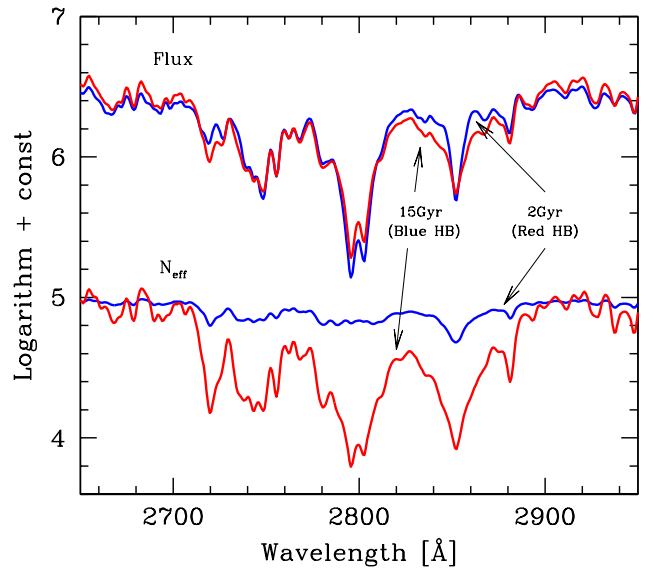}}
\caption[]{Theoretical spectral energy distribution (upper plots) and effective stellar contributors ($N_\mathrm{eff}$) (lower plots) for two SSPs from population synthesis models described by \citet{Buzz89,Buzz93} in the spectral region of the MgII and MgI features around 2800 \AA. The models refer to a 2-Gyr SSP with red HB morphology and a 15-Gyr population with blue HB, as labelled. In spite of the close similarity of the spectra, the two SSPs display a large difference in terms of $N_\mathrm{eff}$, and hence there is much greater statistical scatter in the spectral features expected for the older SSP. Figure from \citet{Buz05}.}
\label{fig:agemet}
\end{figure}

The first implication of the modelling of stellar populations is that the redder the wavelength, the greater is the scatter, since fewer stars contribute to red wavelengths than to blue wavelengths in absolute and relative terms. In fact, each wavelength can fluctuate around the mean of the corresponding distribution of integrated luminosities in a different way, even though it is correlated with the other wavelengths. This naturally implies that for each age and metallicity, each model has its own fitting metrics.

We can take advantage of $N_\mathrm{eff}$ or SBF definitions to theoretically define the weight for each wavelength in a $\chi^2$ fit. In fact, a good $\chi^2$ fit cannot be better than the physical dispersion of the model, which is a physical limit. Exceedance of this physical limit (overfitting) leads to a more precise but erroneous result. Fig.~\ref{fig:metrics}, taken from \citet{CL09}, illustrates the physical dispersion used to identify overfitting for SBFs.

An additional advantage of including the physical weight in the fit is that it breaks degeneracies that are present when observational data are fitted only to the mean value for parametric models. For instance, Fig.~\ref{fig:agemet}, taken from \citet{Buz05}, illustrates breaking of age--metallicity degeneracy using $N_\mathrm{eff}$. Given that the allowed scatter depends on age and metallicity, the resulting $\chi^2$ defines the probability that a fit will produce different results.

Unfortunately, implementation of these ideas is not straightforward. Use of $N_\mathrm{eff}$ directly provides the theoretical weight for a wavelength in a $\chi^2$ fit, but it is dependent on $\cal N$. The use of SBF is independent of $\cal N$. However, it cannot be implemented directly, but requires an iterative process involving a standard fit, use of the SBF to identify overfitted results, and iteration of the process until convergence.

An unexplored area involves taking full advantage of $\rho(L_i,L_j)$, which can be obtained theoretically. In fact, the covariance coefficients and variance define a covariance matrix that can be directly implemented in a $\chi^2$ fit. However, as far as I know, computation of $\rho(L_i,L_j)$ has not been implemented and is not considered in any synthesis code \citep[exceptions are][but they use a Poisson approximation of the stellar luminosity function and the covariance coefficients obtained are not correct]{Cetal01,CVGLMH02,CVG03,GLB04}.

\subsection{The population and the sample definition}

The second implication is related to the definition of the population described by computed distributions. We have seen that we can define populations composed of ${\cal N} = 1$ stars, which are CMD diagrams, and that we can analyse such populations as long as we have enough events $n_\mathrm{sam}$ to describe the distribution.

We can also define populations composed of events with a similar number of stars, $\cal N$, taking advantage of additional information about the source. For instance, we can take the luminosity profile of a galaxy. We can assume that for a given radius, the number of stars is roughly the same (additional information in the form of the geometry of the galaxy is required). Hence, we can evaluate the scatter for the assumed profile, which, independent of observational errors, must be wavelength-dependent. At each galactocentric radius, we are sampling different $\varphi_{\cal N}(L_1\dots L_n)$ distributions with different $n_\mathrm{sam}$ elements. It is even possible that in the outer parts, where $\cal N$ is lower, we find pixels forming a binomial or extremely asymmetric distribution. However, the better our sampling of such distributions, the better will be our characterisation of the population parameters at this radius. Finally, our results for stellar populations must be independent of the radius range chosen once corrected for the radial profile \citep[see][for details on such corrections]{CLJ08}. Hence, we can perform a cross-validation of our results by repeating the analysis by integration over large radial ranges; this means that we reduce $n_\mathrm{sam}$ and increase $\cal N$. The only requirements are: (1) $\cal N$ must be kept constant in each of the elements of the sample ; and (2) sufficient $n_\mathrm{sam}$ elements are required for correct evaluation of the variance (c.f. Eq.~ (\ref{eq:samvarvar})). The results obtained must also be consistent with the stellar population obtained using the integral light for the whole galaxy. Note that if we use SBF we do not need to know the value of $\cal N$, but just need to ensure that it is constant (but unknown) in the $n_\mathrm{sam}$ elements used.

A similar study can be performed using different ways to divide the image. For example, we can take slices of a spherical system and use each slice to compute the variance of the distribution \citep[see][for an example]{Buz05}. A similar technique can be applied to IFU observations. The problem is to obtain $n_\mathrm{sam}$ elements with a similar number of stars and stellar populations that allow us to estimate the scatter (SBF) for comparison with model results. In summary, we can include additional information about the system (geometry, light profile, etc.) in inferences about the stellar populations.

Finally, we can modify the $\varphi_{\cal N}(L_1\dots L_n)$ distributions to include other distributions representing different objects. For instance, the globular cluster distribution of a galaxy \citep[assuming they have the same age and metallicity, in agreement with][]{Yoetal06} implicitly includes a distribution of possible $\cal N$ values. Since these globular clusters have intrinsically low $\cal N$ values, it is possible that some clusters will be in the biased regime described in Fig.~\ref{fig:MonlocationC03}. Since the few clusters dominated by PMS stars are luminous, they will be observed and will be extremely red in colour, even redder than the mean colour of parametric models (Fig.~\ref{fig:Monlocation}). In addition, there would be a blue tail corresponding to faint clusters with low $\cal N$ comprising mainly clusters with low-mass MS stars \citep[see][for more details]{CL04}.

\begin{figure*}
\resizebox{\hsize}{!}{\includegraphics[angle=-90]{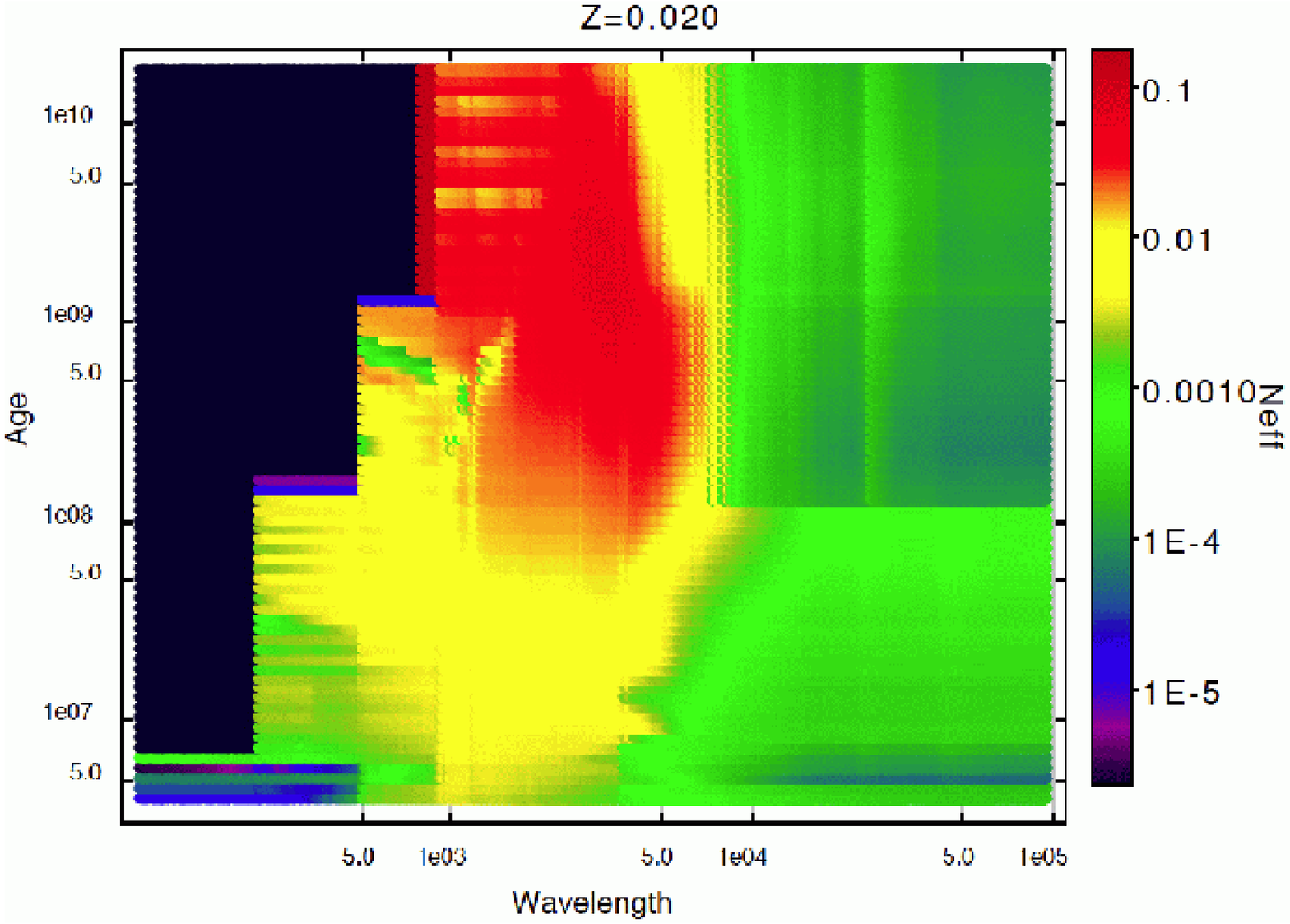}\includegraphics[angle=-90]{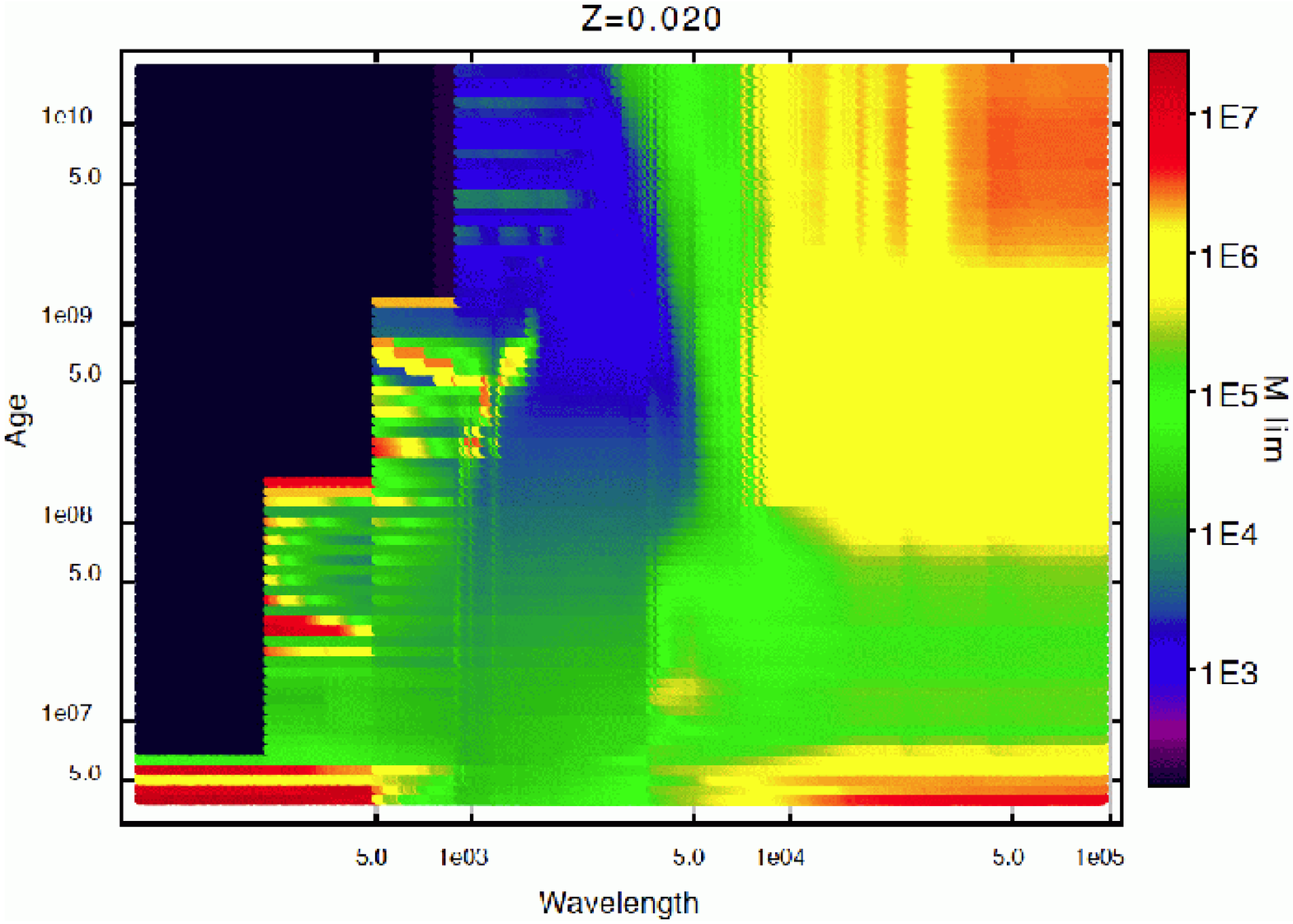}}
\caption[]{Summary of $N_\mathrm{eff}$ values (left) and
mean total mass $M_\mathrm{lim}$ for $\gamma_{1,{\cal N}} \le 0.1$ (right) as a function of L age and L wavelength for $Z=0.020$ SSP models.}
\label{fig:resumen}
\end{figure*}

\subsection{Some rules of thumb}
\label{sec:hurisitcs}

I finish this section with some rules of thumb that can be extracted from the modelling of stellar populations when applied to the inference of physical properties of stellar systems.

First, the relevant quantity in describing possible luminosities is not the total mass of the system, but the total mass (or number of stars) observed {\it for your resolution elements}, $\cal N$. The other relevant quantity is the number of resolution elements $n_\mathrm{sam}$ for a given $\cal N$. The lower the ratio of $\cal N$ to $n_\mathrm{sam}$, the better. In the limit, the optimal case is a CMD analysis.

Second, the scatter depends not only on $\cal N$ but also on the age and wavelength considered. Fig. ~\ref{fig:resumen} shows $N_\mathrm{eff}$ values for SSP models with metallicity $Z=0.020$ for different ages and wavelengths. As I showed earlier, the lower $N_\mathrm{eff}$ is, the greater is the scatter. Fig.~\ref{fig:resumen} shows that blue optical wavelengths with $\lambda < 5000$ \AA ~have intrinsically lower scatter than red wavelengths, independent of $\cal N$. The range 5000--8000 \AA ~has intermediate scatter and scatter increases for wavelengths longer than $\sim 8000$ \AA, depending on the age. In comparison to age determinations that do not consider intrinsic dispersion (i.e. all wavelengths have a similar weight), the safest age inferences correspond to ages between 8 and $\sim$ 200 Myr since variation of $N_\mathrm{eff}$ values with wavelength is lower in this mass range.

Fig.~\ref{fig:resumen} also shows the mean total mass for $\gamma_{1,{\cal N}} \le 0.1$ as a function of age and wavelength for $Z=0.020$ SSP models. This is a quantitative visualization of how the average total mass needed to reach a Gaussian-like regime varies with age and wavelength. It is evident that blue optical wavelengths reach a Gaussian regime at a lower average total mass when compared to other wavelengths. In contrast, red wavelengths not only have greater scatter, but this scatter is also associated with non-Gaussian distributions for a wider range of average total mass. Thus, we can strengthen the statement about age inferences: the safest age inferences correspond to ages between 8 and $\sim$ 200 Myr if the cluster mass is greater than $10^5 \mathrm{M}_\odot$.

Third, the input distributions ${\cal B}(m_0,t,Z)$ or IMF and SFH define the output distributions. It should be possible to obtain better fits (more precise, but not necessarily more accurate) by changing the input distributions. However, we must be aware that we have explored the possible output distributions before any such changes.

For instance, low-mass clusters have strong fluctuations around the IMF; each cluster (each IMF realization) could produce an excess or deficit of massive stars. Using the mean value obtained by parametric models, a top-heavy or bottom-heavy IMF would produce a better fit of models and data. Such IMF variations would undoubtedly be linked to variations in age and the total mass/number of stars for a system. However, Monte Carlo simulations can also produce a better fit without invoking any IMF variation. It is possible that Monte Carlo simulations using distributions of different IMFs (e.g. combining IMFs with a variable lower or upper mass limit with a distribution of possible lower or upper mass limits) would produce even better fits. I am sure that the approach using only the mean value is methodologically erroneous. However, it is not clear which of the two solutions obtained by Monte Carlo simulations is the best unless one of the hypotheses (fixed IMF or a distribution combining different IMFs) is incompatible with observational data. 

As a practical rule, before exploring or claiming variations in the input parameters, a check is required to ensure that such input parameters are actually incompatible with observational data.

Another issue is how to evaluate scatter outside the SSP hypothesis. Formally, we may consider any SFH as a combination of SSPs. Hence, for any SFH scenario evaluated at time $t_\mathrm{mod}$, we can assume that the scatter can be evaluated using the most restrictive SSP situation in the time range from 0 to $t_\mathrm{mod}$. For instance, Fig.~\ref{fig:resumen} shows that the ionising flux ($\lambda \lesssim 912$ \AA) requires an average total mass greater than $10^5 \mathrm{M}_\odot$ to reach a Gaussian-like regime. Hence, we must ensure that at least $10^5 \mathrm{M}_\odot$ has been formed in each SSP comprising the SFH. Assume that 1 Myr is the time interval used to define a star formation rate. This implies that there would be no Gaussian-like distributions for SFR less than $0.1 \mathrm{M}_\odot \mathrm{year}^{-1}$ (i.e. there would be a bias in the inferences obtained using the mean obtained by parametric models). However, a more quantitative study of this subject is required. \citet{slug} have suggested additional ideas on the evaluation of scatter including the SFH.

Fourth, regarding the output parameters and inferences for time and the total mass/number of stars, we can summarize the following rules:

\begin{enumerate}

\item {\it Use all available information for the system, including previous inferences}. However, wavelength ranges used for inferences in the literature must be considered. Additional criteria, such as those in the following points, can be used to evaluate roughly which inferences are more reliable.

Additional information on the system can be obtained from images and other data that, although not used directly in the inferences, constrain the possible range of solutions. Recovery of a complete picture of the system compatible with all the available information should be the aim, and not just a partial picture that can be drawn from particular data.

It is particularly useful to look for the `smoking guns' for age inference: for example, emission lines in star-forming systems imply an age less than 10 Myr ; Wolf--Rayet stars imply an age less than 6 Myr (neglecting binary systems); supernova emission or supernova remnants (from optical, radio or X-ray observations) imply an age less than 50 Myr; high-mass X-ray binaries imply previous supernova events, and hence an age greater than 3 Myr. For instance, \citet{FLCW12} showed that the use of broad-band photometry with narrow-band H$_\alpha$ photometry greatly improves the quality of inferences. However, note that the presence of a `smoking gun' helps to define age ranges, but the absence of smoking guns does not provide information if the mass/number of stars in observations is not known.

\item {\it Always obtain an estimate of the mass/number of stars in the resolution element}. The confidence of an age inference cannot be evaluated unless an estimate of the mass/number of stars of such an age has been obtained \citep[see][for additional implications of this point]{masscleannage}.

\item {\it Identify the integrated luminosity distribution regime for the system considered }. $\chi^2$ fitting including the physical variance and covariance coefficients is optimal for the Gaussian regime, but it fails for other distributions. Fig.~\ref{fig:resumen} can be used to identify Gaussianity. For large wavelength coverage, for which different regimes would be present, rejection of some parts of the spectra in the fit can be considered; it is better to obtain a less precise but more accurate result than a very precise but erroneous result caused by overfitting; in any case, such information can be used as a guide to obtain a complete picture of the system.

\item {\it When using diagnostic diagrams (indices), compare the location of the parametric model, individual stars and observations.} It is especially useful to identify the origin of outliers in diagnostic diagrams, and to evaluate the expected range of scatter in the model.

\item In general, {\it CMDs analyses are more robust than analyses of integrated spectra}. If such information is available, use it and do not try surpass CMDs, which is simply impossible. The sum of (unknown) elements in a sample cannot provide more information than knowledge for all the particular elements of the sample.

\item {\it Blue wavelengths (3000--5000 \AA) are robust}. We have shown that blue wavelengths have intrinciscally lower physical scatter than any other wavelengths. Hence, when comparing different inferences in the literature, those based on just blue wavelengths are the most accurate. They may not be the most precise, but blue wavelengths always provide better fitting than any other wavelength. Hence, take the greatest possible advantage of blue wavelengths (e.g. when using normalized spectra to a given wavelength to obtain SFH, use blue wavelengths).

\item {\it A good solution is the distribution of possible solutions}. I discussed this in Section \ref{sec:metrics}, but I would like to emphasise the point. The best $\chi^2$ value would be a numerical artefact (e.g. local numerical fluctuations). For instance, it seems surprising that codes that infer the SFH using the mean value of parametric models do not usually quote any age--metallicity degeneracy, although it is present at a spectral level (c.f. Fig.~\ref{fig:agemet}). In fact, it is an artefact of using only the best $\chi^2$ fit when results are presented. Again, I refer to \citet{FLCW12} as an example of how the use of the distribution of possible solutions improve inferences.
\end{enumerate}

\section{The (still) poorly explored arena}
\label{sec:open}

In the preceding sections I raised several questions on the application of population synthesis models, ranging from their use in CMD diagrams for semi-resolved systems to fitting metrics and the still unexplored arena of covariance coefficients. In fact, most of the applications presented here have not been fully developed and have been presented in conference papers. In this respect I refer to work by \citet{Buz05}, which has been a source of inspiration to most of the applications described here and some additional applications not discussed. However, let me finish with two comments about additional unexplored areas in which stochasticity could play a role.

The first is chemical evolution. At the start of this paper I noted that any modelling that makes use of the stellar birth rate (IMF and SFH) is intrinsically probabilistic. Hence, the present discussion also applies to chemical evolution models. Here, there is no stellar luminosity function, but a stellar yield function, which is the amount of material that a star has ejected (instantaneously or accumulatively). Obviously, such a stellar yield function refers mainly to dead stars, stars with $\ell =0$ in the stellar luminosity function, and is hence correlated with live stars. The application of such ideas is suggestive, but not simple. Chemical evolution affects the metallicity and depends on the SFH; hence, metallicity and SFH cannot be separable functions in the stellar birth rate. Even worse, the actual metallicity of a system depends on its previous metallicity evolution, which is itself described as a probability distribution. Determination of the evolution of a system would require a change in the ordinary differential equations defining chemical evolution to stochastic differential equations. Interested readers can find very preliminary approaches to the subject in work by \citet{Ceretal00,Cetal01,CM02}, and more detailed studies by \citet{SF95,CH08}.

The second question is the inclusion of highly variable phases such as thermal pulses in synthesis models. The present idea is actually a re-elaboration of the so-called fuel consumption theorem proposed by \citet{RB83} and \citet{Buzz89} applied over isochrone synthesis. A requirement of isochrone synthesis is that evolutionary tracks must be smooth enough to allow interpolations and compute isochrones. In particular, variation is not allowed for a star with a given mass and a given age. However, we can include variability as far as we can model it by a probability distribution function. In fact, we can define such a probability as proportional to the period of variation and include it directly in the stellar luminosity function. I refer readers interested in this subject to the paper by \citet{GGS04}, whose ideas provide a formal development of the probabilistic modelling of stellar populations.

\section{Conclusions}
\label{sec:conclu}

We studied the modelling of stochasticity for stellar populations. Such stochasticity is intrinsic in the modelling in that we do not directly know the individual stars in a {\it generic} stellar system. We must use probability distributions and assign probabilities for the presence of each individual star. The only possible result of the modelling is a probabilistic description including {\it all} possible situations.

The input probability distribution is the stellar birth rate ${\cal B}(m_0,t,Z)$, which represents the probability that an individual star was born with a given initial mass $m_0$ at a given time $t$ with a given metallicity $Z$. This probability distribution is modified by the transformations provided by stellar evolution theory and evaluated at different times $t_\mathrm{mod}$. The resulting distribution, the stellar luminosity function, represents the probability that a star had a given luminosity set (e.g. a set of colours or wavelengths), $\ell_1 \dots \ell_n$, at a given time.

The probability distribution of the integrated luminosity of a system with $\cal N$ stars can be obtained directly from the stellar luminosity function in an exact way via a self-convolution process. It can also be obtained from a set of $n_\mathrm{sim}$ Monte Carlo simulations. Finally, it can be described as a function of parameters of the distribution (mean, variance, skewness, and kurtosis) that is related to the stellar luminosity function by simple scale relations. Standard synthesis models use a parametric description, although, with some exceptions, they only compute the mean value of the distribution.

Hence, the results of standard (parametric) modelling can be used in any situation, including ${\cal N} =1$ (i.e. CMDs), as long as we understand that it is only a mean value of possible luminosities. However, mean values are not useful unless we know the shape of the distribution. In the most optimistic case, for large $\cal N$ values, the distribution is Gaussian and is hence defined by the mean and variance of each luminosity, and the covariance between the different luminosities. The value of $\cal N$ when this Gaussian regime is reached can be obtained by analysis of the skewness and kurtosis of the stellar luminosity function.

Monte Carlo modelling (or the self-convolution process) is a useful approach outside the Gaussian regime. However, Monte Carlo simulations can only be fully understood if they are analysed with a parametric description of the possible probability distributions. In addition, Monte Carlo simulations are an ideal tool for including constraints in fine-tuned modelling of particular objects or situations. A drawback is that large sets of Monte Carlo simulations implicitly include additional priors, such as the distribution of ages or $\cal N$ values. Such priors must be chosen carefully in any application inferring physical parameters from observations since possible inferences depend on them according to the Bayes theorem. Finally, it is not possible to evaluate the reliability of Monte Carlo simulations without explicit knowledge of the number of simulations $n_\mathrm{sim}$, $\cal N$, and all the assumed priors.

The fact that even in the most optimistic case of ${\cal N} = \infty$, the distribution of integrated luminosities is Gaussian (where the parameters of the Gaussian distribution depend on size, age, and metallicity) has major implications for the use of synthesis models to obtain inferences. The principal one is that each model and wavelength has its own physical scatter. This defines a metric of fitting for each model (i.e. not all wavelengths are equivalent in the fit and they must be weighted according their own physical scatter). An advantage of the use of such fitting metrics is that it breaks degeneracies in physical properties, such as age--metallicity degeneracy. A second implication is that we can obtain physical information from observed scatter.

Finally, throughout the paper we explored additional implications and applications of the probabilistic modelling of stellar populations. Most of them are only tentative ideas, but they increase the predictive power of synthesis models and provide more accurate inferences of physical parameters from observational data.

\vspace{0.5cm}

{\it Addendum after publication (only in astro-ph version):} After the publication of this review, appear in the literature an additional paper dealing with modeling stellar populations by means of Monte Carlo simulations by \cite{Andersetal13}. Models results and programs related with that work will  available at: {\tt  http://data.galev.org/models/anders13}.

\section{Acknowledgements}

My work is supported by MICINN (Spain) through AYA2010-15081 and AYA2010-15196 programmes. I acknowledge the development of TopCat software \citep{TopCat} since it helped me greatly in obtaining Figs. \ref{fig:mon1} and \ref{fig:resumen}. This review compiles the efforts of many years of thinking and working with stellar population models and stochasticity, and it would not have been possible without the collaboration of Valentina Luridiana. I also acknowledge Alberto Buzzoni and Steve Shore for suggestions, comments and discussions on the modelling and applications of these ideas, most of then outlined here. I acknowledge Roberto and Elena Terlevich, David Valls Gabaud, Sandro Bressan, Mercedes Molla, Jes\'us Ma{\'{\i}}z Apellaniz, and Angel Bongiovani for useful discussion. Also, I acknowledge Carma Gallart and J. Ma{\'{\i}}z Appelaniz for providing me some figures for this review, and Sally Oey for her feedback and to give me the opportunity of write this review. Finally, I acknowledge Nuno and Carlos (twins), Dario and Eva who show experimentally that Nature fluctuates around theoretical expectations (no-twins), and that such fluctuations make life (and science) more interesting.

\bibliographystyle{elsarticle-harv}

\end{document}